\documentclass[10pt,aps,prl,twocolumn,notitlepage,showpacs,superscriptaddress]{revtex4-1} 
\usepackage{amsthm,amsmath,amsfonts,amssymb,verbatim,color} 
\usepackage{graphicx} 
\usepackage{subfigure}
\usepackage{bm}
\usepackage{epsfig,slashed}        
\usepackage[T1]{fontenc}
\usepackage[colorlinks=true,citecolor=blue,linkcolor=blue,urlcolor=blue]{hyperref}
\begin{document}

\newcommand{\tr}{\mathop{\mathrm{Tr}}}
\newcommand{\bsigma}{\boldsymbol{\sigma}} 
\newcommand{\re}{\mathop{\mathrm{Re}}}
\newcommand{\im}{\mathop{\mathrm{Im}}}
\renewcommand{\b}[1]{{\boldsymbol{#1}}}  
\newcommand{\diag}{\mathrm{diag}}
\newcommand{\sign}{\mathrm{sign}}
\newcommand{\sgn}{\mathop{\mathrm{sgn}}}
\renewcommand{\c}[1]{\mathcal{#1}}  
\renewcommand{\d}{\text{\dj}}

\newcommand{\mb}{\bm}
\newcommand{\ua}{\uparrow}
\newcommand{\da}{\downarrow}
\newcommand{\ra}{\rightarrow}
\newcommand{\la}{\leftarrow}
\newcommand{\mc}{\mathcal}
\newcommand{\bs}{\boldsymbol}
\newcommand{\lra}{\leftrightarrow}
\newcommand{\half}{{\textstyle{\frac{1}{2}}}}
\newcommand{\mf}{\mathfrak}
\newcommand{\MF}{\text{MF}}
\newcommand{\IR}{\text{IR}}
\newcommand{\UV}{\text{UV}}

\newcommand{\labell}[1]{\label{#1}}
\newcommand{\etal}{{\it et.\ al. }}
\newcommand{\bea}{\begin{eqnarray}}
\newcommand{\eea}{\end{eqnarray}}
\newcommand{\ba}{\begin{eqnarray}}
\newcommand{\ea}{\end{eqnarray}}
\newcommand{\nn}{\nonumber \\}
\newcommand{\eqn}[1]{(\ref{#1})}
\newcommand{\beq}{\begin{equation}}
\newcommand{\eeq}{\end{equation}}
\newcommand{\beqa}{\begin{eqnarray}}
\newcommand{\eeqa}{\end{eqnarray}}
\newcommand{\beqar}{\begin{eqnarray*}}
\newcommand{\eeqar}{\end{eqnarray*}}
\newcommand{\e}{\epsilon}
\newcommand{\reef}[1]{(\ref{#1})}
\newcommand{\ssc}{\scriptscriptstyle}
\newcommand{\eg}{{\it e.g.,}\ }
\newcommand{\ie}{{\it i.e.,}\ }
\renewcommand{\comment}[1]{{\bf [[[#1]]]}}
\newcommand{\equ}[1]{(\ref{#1})}
\newcommand{\mt}[1]{\textrm{\tiny #1}}
\newcommand{\ang}[1]{\left\langle #1 \right\rangle}
\newcommand{\veps}{\varepsilon}
\newcommand{\X}{\mathcal{X}}
\newcommand{\W}{\mathcal{W}}
\newcommand{\R}{\mathcal{R}}
\newcommand{\Z}{\mathcal{Z}}
\newcommand{\A}{\mathcal{A}}
\newcommand{\B}{\mathcal{B}}
\newcommand{\C}{\mathcal{C}}
\newcommand{\D}{\mathcal{D}}
\newcommand{\E}{\mathcal{E}}
\newcommand{\G}{\mathcal{G}}
\newcommand{\cO}{\mathcal{O}}
\newcommand{\cM}{\mathcal{M}}
\newcommand{\hH}{\mathcal{H}} 
\newcommand{\nvec}{{\vec{n}}}
\newcommand{\lgb}{\lambda_{\text{\tiny{GB}}}}
\newcommand{\tL}{\tilde{L}}
\newcommand{\nc}{N_c}
\newcommand{\pd}{\partial}
\renewcommand{\la}{\lambda}
\newcommand{\lp}{\ell_{\mt P}}
\newcommand{\mbf}{\mathbf}
\newcommand{\fin}{f_\infty}
\newcommand{\ff}{f_{\infty}}
\newcommand{\ct}{C_{T}} 
\newcommand{\hi}{{\hat \imath}}
\newcommand{\hj}{{\hat \jmath}}
\newcommand{\qn}{\textswab{q}}
\newcommand{\wn}{\textswab{w}}
\newcommand{\nw}{N_W}
\newcommand{\hr}{{\hat r}}
\newcommand{\vrho}{\varrho}
\newcommand{\trho}{{\tilde\varrho}}
\newcommand{\ttau}{{\tilde \tau}}
\newcommand{\ta}{{\tilde a}}
\newcommand{\tb}{{\tilde b}}
\newcommand{\tc}{{\tilde c}}

\newcommand{\ads}{a_d^*}
\newcommand{\pe}{\rho_\mt{E}}
\newcommand{\hC}{{\widehat C}}
\newcommand{\te}{t_\mt{E}}
\newcommand{\too}[1]{\mathrel{\mathop \to_{\scriptscriptstyle{#1}}}{}\!\!}
\newcommand{\dr}{{\Delta R}}
\newcommand{\m}{\text{min}}
\renewcommand{\c}{$c$}
\newcommand{\F}{$F$}
\newcommand{\cA}{{\cal A}}
\newcommand{\mv}{{\cal a}}
\newcommand{\um}{{\widehat{\bf m}}}
\newcommand{\Sig}{{\partial V}}
\newcommand{\un}{{\hat{\bf n}}}
\newcommand{\ut}{{\hat{\bf t}}}
\newcommand{\uu}{{\hat{\bf u}}}

\newcommand{\ren}{R\'enyi\ }
\newcommand{\sg}{\sigma}
\newcommand{\q}{a}
\newcommand{\qe}{\q_{\rm\ssc E}}
\newcommand{\see}{S} 
\newcommand{\ka}{\kappa}
\newcommand{\kae}{\ka_{\rm \ssc E}}
\newcommand{\sge}{\sg_{\rm \ssc E}}
\newcommand{\cs}{c_{\ssc S}}
\newcommand{\cse}{c_{\ssc S,E}}
\newcommand{\ctt}{C_{\ssc T}}
\newcommand{\ctte}{C_{\ssc T\!, {\rm E}}} 
\newcommand{\cttgb}{C_{\ssc T,GB}}
\newcommand{\lhat}{\hat\lambda}
\newcommand{\alhat}{\hat\alpha}

\renewcommand{\mc}{\mathcal}
\newcommand{\al}{\alpha}
\newcommand{\de}{\delta}
\newcommand{\ts}{\thinspace{}}
\newcommand{\req}[1]{(\ref{#1})} 

\newcommand{\red}[1]{{\color{red}#1}}

\newcommand\Ts{\rule{0pt}{4.1ex}}       
\newcommand\Bs{\rule[-2.8ex]{0pt}{0pt}} 

\DeclareGraphicsExtensions{.png}      

\title{Optical conductivity of topological surface states \\ with emergent supersymmetry}    

\author{William Witczak-Krempa} 
\affiliation{Department of Physics, Harvard University, Cambridge, Massachusetts 02138, USA}

\author{Joseph Maciejko}
\affiliation{Department of Physics, University of Alberta, Edmonton, Alberta T6G 2E1, Canada}
\affiliation{Theoretical Physics Institute, University of Alberta, Edmonton, Alberta T6G 2E1, Canada}
\affiliation{Canadian Institute for Advanced Research, Toronto, Ontario M5G 1Z8, Canada}

\date\today

\begin{abstract}

Topological states of electrons present new avenues to explore the rich phenomenology of correlated quantum matter.
Topological insulators (TIs) in particular offer an experimental setting to study novel quantum critical points (QCPs)
of massless Dirac fermions, which exist on the sample's surface. 
Here, we obtain exact results for the zero- and finite-temperature optical conductivity at 
the semimetal-superconductor QCP for these topological surface states.
This strongly interacting QCP is described by a scale invariant theory with 
emergent supersymmetry, which is a unique symmetry mixing bosons and fermions. We show that supersymmetry implies exact relations between the optical conductivity and two otherwise unrelated properties: the shear viscosity and the entanglement entropy. We discuss experimental considerations for the observation of these signatures in TIs.
\end{abstract}  

\pacs{
05.30.Rt,		
74.40.Kb,	
71.27.+a,	
11.30.Pb		
}

\maketitle

 
\begin{figure}
\center
   \includegraphics[scale=.29]{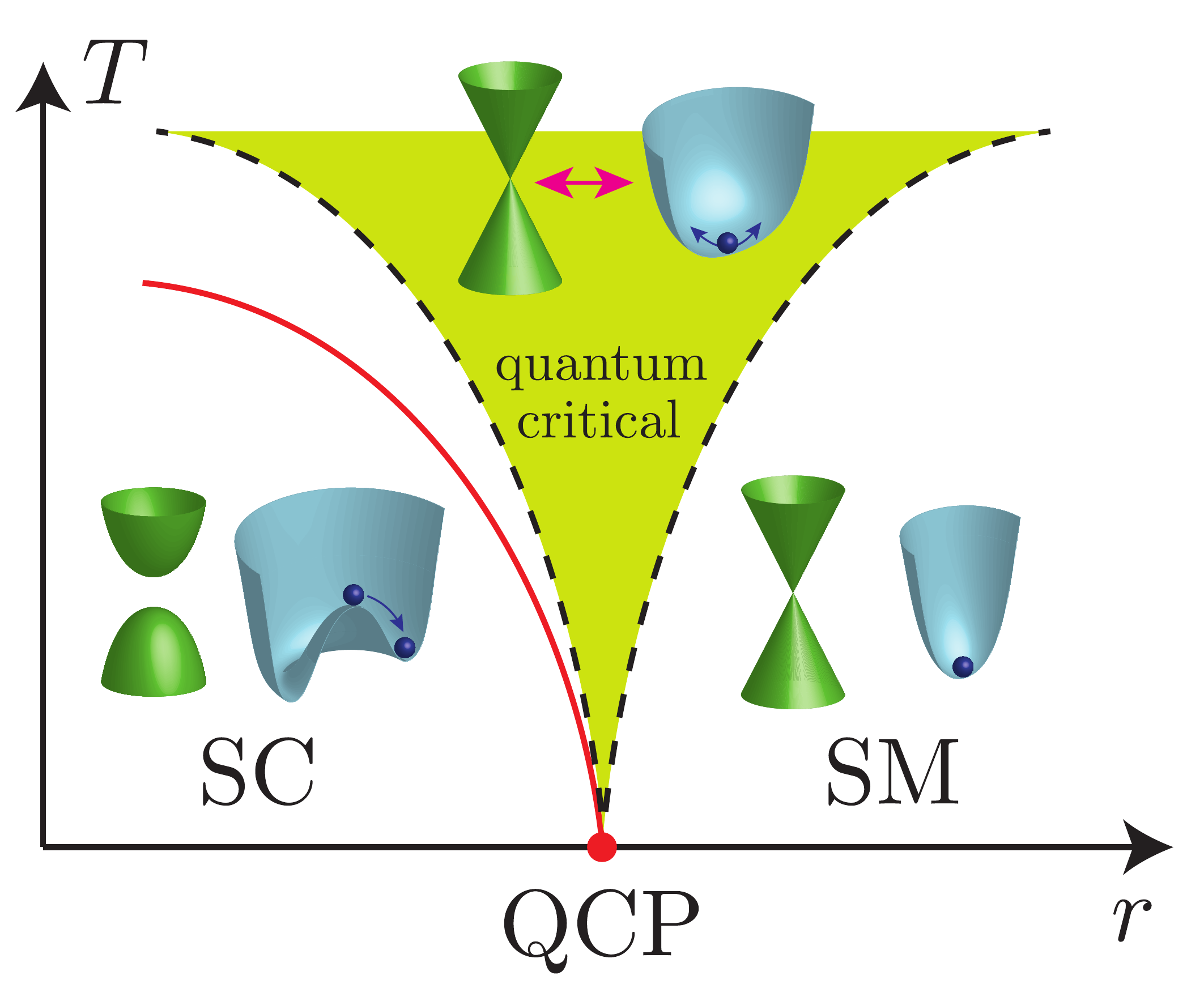}     
   \caption{Phase diagram near the semimetal-superconductor (SM-SC) quantum critical point of Dirac fermions
on the surface of a 3D topological insulator. $T$ is the temperature and $r$ is the nonthermal tuning parameter [see Eq.~(\ref{L})]. The evolution of the Dirac dispersion and Cooper field potential are shown. Supersymmetry emerges at the QCP where it relates the Dirac fermions and the bosonic Cooper pairs. 
} 
\label{fig:corner}
\centering        
\end{figure}     
Topological insulators~\cite{RMP_TI,RMP_TI2} allow for the experimental study of new quantum states of matter. The strong spin-orbit 
coupling in these bulk-insulating materials leads to unique gapless Dirac fermion surface states. These can undergo quantum phase transitions
forbidden in non-topological systems, and thus constitute a new platform to study the rich physics of quantum criticality~\cite{cenke1,*cenke2,lundgren}. 
A considerable challenge in the study of interacting QCPs is to determine their dynamical response---that is, their response at finite frequency $\omega$---both at zero and  
finite temperature $T$, such as the optical conductivity $\sigma(\omega,T)$.    
Here, we focus on the dynamical response of a novel QCP that can appear at the surface of a three-dimensional (3D) topological insulator:
it describes the interaction-driven quantum phase transition between a single Dirac cone of electrons and a gapped 
superconductor~\cite{grover2014,ponte2014} (see Fig.~\ref{fig:corner}). As an important step
towards observing this transition, recent experiments have reported the discovery of intrinsic   
superconductivity on the surface of a 3D topological insulator, Sb$_2$Te$_3$~\cite{zhao2015}.   
We emphasize that standard 2D (or layered) systems that do not break time-reversal symmetry must have an even number of Dirac cones and thus cannot host
this transition. More complex scenarios realizing multiple copies of this QCP can occur 
via $f$-wave pairing~\cite{lee2007} and pair-density-wave~\cite{jian2015} instabilities of spinless Dirac 
fermions on the 2D honeycomb lattice, or for interacting ultracold atomic gases in optical lattices~\cite{yu2010}. 
   

\begin{table*} 
  \centering
  \begin{tabular}{c||c|c}   
     & {\bf Dirac SM-SC}  & SC-Insulator \\
    \hline \hline 
    $\sigma_\infty$ & $\; \dfrac{5(16\pi-9\sqrt{3})}{243\pi} \approx 0.227 \; $   &  $0.226$ \Ts\Bs \\ \hline 
    $\eta_\infty$ & $\sigma_\infty/40\approx 5.68\times 10^{-3}$ & $\;3.68\times 10^{-3}$ \Ts\Bs \\ \hline  
    $\lambda_{\rm corner}$ & $ \sigma_\infty/20 \approx 0.0113$ & $0.00737$ \Ts\Bs \\ \hline 
    $b'$ & 0 &  $-0.3(1)$ 
  \end{tabular} 
  \caption{{\bf Exact results.} Comparison of the conductivity, viscosity, and entanglement entropy at two different QCPs. Left column: exact results obtained in this paper for the Dirac semimetal (SM) to superconductor (SC) QCP with
emergent supersymmetry. Right column: known approximate results for the SC to Cooper-pair-insulator QCP.  
  The optical conductivity and dynamical shear viscosity at $T=0$ 
  are $\sigma(\omega,0)=\sigma_\infty e^2/\hbar$ and $\eta(\omega,0)=\eta_\infty\omega^2 \hbar$. 
$\lambda_{\rm corner}$ determines the entanglement entropy of nearly smooth corners [Eq.~(\ref{atheta})]. $b'$ determines a finite-$T$ correction
to the optical conductivity 
of the form $b'\,(ik_B T/\hbar\omega)^3$.     
\label{tbl:ratio}} 
\end{table*}       

When the chemical potential is at the Dirac point, a special type of symmetry emerges at the 
QCP~\cite{lee2007,grover2014,ponte2014,jian2015,ScottThomas}: spacetime supersymmetry (SUSY).
SUSY relates bosons and fermions, and has been proposed to exist in extensions of the Standard Model of elementary particle
physics, but has not yet been observed. At the QCP of Fig.~\ref{fig:corner}, it emerges naturally by relating 
the Dirac fermions of the semimetal to the bosonic Cooper pairs of the superconductor. These two become
degenerate at the transition and in fact share a deeper relation described by SUSY. We emphasize that this is a consequence of the strong
interactions at the QCP, where long-lived excitations (quasiparticles) are destroyed by quantum zero-point fluctuations.  
We show that even in the presence of such strong interactions, SUSY allows the \emph{exact}  
determination of the zero-temperature optical conductivity $\sigma(\omega,0)$ of the topological surface states at the QCP. We are not aware of any known exact result for the dynamical response of a realistic strongly interacting QCP in spatial dimensions higher than one. In addition, SUSY implies that the conductivity directly determines the shear viscosity and certain many-body entanglement properties. Our exact findings are summarized in Table~\ref{tbl:ratio}.   
We begin by describing the low-energy theory of the QCP, and then explain how the emergent SUSY allows the exact
determination of various properties such as the optical conductivity. We end by discussing considerations relevant for the experimental 
observation of these signatures. 
  
The generalized Landau-Ginzburg theory for the quantum phase transition couples a single charge-$e$ Dirac fermion $\psi$ to the charge-$2e$ Cooper pair   
bosonic field, $\phi$,   
\begin{multline} \label{L}
  \mathcal L = i\bar\psi\gamma_\mu \partial_\mu \psi+\frac{1}{2}|\partial_\mu\phi|^2+\frac{r}{2}|\phi|^2 
  +\frac{\lambda}{2}|\phi|^4 \\ + h(\phi^*\psi^Ti\gamma_2\psi+{\rm c.c.}), 
\end{multline} 
in imaginary time, where $\bar\psi=\psi^\dagger\gamma_0$ and $\gamma_\mu$, $\mu=0,1,2$ are $2\times 2$ matrices satisfying the Pauli algebra. We note that time-reversal invariance forbids a fermion mass term.
The QCP is obtained by tuning $r$ to zero, and the resulting system is strongly correlated because both the quartic coupling $\lambda$ and the fermion-boson 
coupling $h$ are relevant at the noninteracting, UV fixed point $\lambda=h=0$. There is a single 
stable IR fixed point with $\lambda=h^2\neq 0$~\cite{grover2014,ponte2014,lee2007,jian2015,ScottThomas}, at which (\ref{L}) 
becomes invariant under SUSY transformations that rotate the Dirac fermion into the boson and vice-versa~\cite{aharony1997}. 
In line with the requirement of SUSY, it was shown \cite{lee2007,grover2014} that the fermion and Cooper pair velocities flow to the same value at low energies, which  
we henceforth set to unity. This is consistent with the fermions and Cooper pair fields being strongly coupled.
As a result of the unique velocity, (\ref{L}) displays emergent Lorentz invariance.    
By virtue of SUSY, the fermion and boson anomalous dimensions are known \emph{exactly}~\cite{aharony1997}:    
$\eta_\psi=\eta_\phi=1/3$, a clear indication of the destruction
of quasiparticles. The electric current is given by the sum of fermionic and bosonic contributions: $J_\mu=\bar{\psi}\gamma_\mu\psi+i(\phi^*\partial_\mu\phi-\mathrm{c.c.})$.    

The QCP (\ref{L}) has an important purely bosonic analog obtained by omitting the fermions, 
in which case it describes the superconductor-to-insulator quantum phase transition obtained by localizing Cooper pairs~\cite{fisher1990}. 
Part of the interest in this QCP (and its optical conductivity)  
comes from the fact that it is believed to occur in certain thin-film superconductors~\cite{fisher1990}.   
The QCP that we study belongs to a different universality class because it involves fermions, and we shall contrast 
the two throughout (see Table~\ref{tbl:ratio}).      
  
{\bf Exact charge \& shear responses}:  
As the system is tuned to the QCP, the optical conductivity depends only on the ratio $\hbar\omega/k_BT$~\cite{damle1997}: 
\begin{align}  
  \sigma(\omega,T) = \frac{e^2}{\hbar}\, \Phi\left(\frac{\hbar\omega}{k_BT}\right),    
\end{align}
where $\Phi(x)$ is a dimensionless, universal scaling function that is fully determined by the universality class of the transition.  
We recall that the conductivity is obtained from the current-current correlator via the Kubo formula,
$\sigma \!=\! \frac{1}{i\omega}\langle J_x(\omega, \vec k\!=\!0) J_x(-\omega, \vec k\!=\!0) \rangle_T$.
An important consequence of the scale invariance is that the optical conductivity at $T\!=\!0$
is a frequency-independent \emph{constant}: $\sigma(\omega,0)\!=\!e^2\,\sigma_\infty/\hbar$,  
where we have defined $\sigma_\infty= \Phi(\infty)$, and we are working at frequencies lesser than 
microscopic energy scales such that we are probing the universal response.   
For QCPs such as the one under consideration,
this universal constant determines the charge response of the ground state in a system lacking quasiparticles. We now
describe how the emergent SUSY can be used to compute $\sigma_\infty$ exactly.     

In supersymmetric field theories, operators are organized into representations of the SUSY algebra called supermultiplets, 
the same way spin operators are organized into representations of $SU(2)$.  
In our case, the electric current $J_\mu$ lies in the same supermultiplet as the stress tensor $T_{\mu\nu}$, 
the so-called supercurrent supermultiplet~\cite{closset2013}. Here, supercurrent does not refer to superconductivity but rather to the Noether current associated with SUSY. 
One associates to each supermultiplet a so-called superfield which contains all the various components of the supermultiplet. The superfield associated with the supercurrent supermultiplet is denoted $\mathcal{J}_\mu$, and is highly constrained by SUSY. 
Crucially, the two-point correlation function of the  
supercurrent is entirely fixed up to an overall multiplicative constant~\cite{osborn1999,barnes2005,buchbinder2015}, denoted $\mathcal C$. 
Because $\mathcal{J}_\mu$ contains both the current $J_\mu$ and the stress tensor $T_{\mu\nu}$, 
this implies a relation between their respective two-point correlation functions.   
This relation in turn implies a nontrivial relation between the universal charge and shear responses at the QCP (\ref{L}).
  
In 2D QCPs with emergent Lorentz invariance, the two-point correlation functions of the   
current and the stress tensor have the power-law forms~\cite{osborn1994} 
$\langle J_\mu(x)J_\nu(0)\rangle \!=\!C_J\frac{I_{\mu\nu}(x)}{|x|^4}$ and $\langle T_{\mu\nu}(x)T_{\rho\sigma}(0)\rangle \!=\!C_T\frac{I_{\mu\nu,\rho\sigma}(x)}{|x|^6}$, where $x$ denotes the spacetime separation, the $I$'s are dimensionless tensors without free parameters~\cite{SuppMat},\vphantom{\cite{dumitrescu2011,hofman08,myers2011,chowdhury2013}} and the constants $C_{J,T}$ are universal low-energy properties    
related to the conductivity and viscosity, respectively, as we shall see below. The above discussion implies that these are both  
proportional to $\mathcal C$ in our SUSY QCP, 
hence their ratio is fixed. We find that the particular SUSY of (\ref{L}) imposes $C_J/C_T = 5/3$~\cite{SuppMat}. 
This then leads to a universal ratio between the zero-temperature dynamical shear  
viscosity and optical conductivity at the strongly interacting QCP (\ref{L}).   
The dynamical shear viscosity $\eta(\omega,T)$ is given by the two-point function of the $xy$-component of the stress  
tensor~\cite{bradlyn2012,witczak-krempa2015}, and becomes $\eta(\omega,0) =\eta_\infty\, \omega^2 \hbar$ at zero 
temperature. 
Fourier transforming from time to frequency, we find $\sigma_\infty=\pi^2 C_J/2$ and $\eta_\infty=\pi^2 C_T/48$, and thus the universal ratio
\begin{align} \label{sig-eta}
\frac{\sigma_\infty}{\eta_\infty}=40,
\end{align}
which is a nontrivial fingerprint of the emergent SUSY at the QCP of (\ref{L}). 
We emphasize that in the absence of SUSY no relation exists in general between the conductivity and 
shear viscosity of QCPs. In fact, (\ref{sig-eta}) is violated at the superconductor-insulator QCP of Cooper pairs, see Table~\ref{tbl:ratio}.  \\ 

The emergent SUSY allows the exact calculation of the optical conductivity by geometric methods, and 
crucially relies on the connection between the conductivity and viscosity (\ref{sig-eta}).  
First, the shear viscosity coefficient $\eta_\infty$, or alternatively $C_T$, can be obtained from the second derivative of the 
free energy on the squashed three-sphere with respect to the squashing parameter~\cite{closset2013}. 
Remarkably, the free energy on this spacetime geometry can be computed exactly using the so-called SUSY localization 
technique~\cite{imamura2011,imamura2012}, even if the theory is 
strongly coupled. For the QCP of interest to us, an integral expression for $C_T$ was recently 
obtained~\cite{nishioka2013}, which can be computed numerically.  
We were able to evaluate this integral in closed form~\cite{SuppMat}. We then used the relation (\ref{sig-eta}) to obtain 
an exact result for the $T=0$ optical conductivity at the  
semimetal-superconductor QCP of 2D Dirac fermions:   
\begin{align}\label{SigmaInfWZ} 
\sigma(\omega,0)= \frac{5(16\pi-9\sqrt{3})}{243\pi}\, \frac{e^2}{\hbar} \approx 0.2271\, \frac{e^2}{\hbar}. 
\end{align} 
To our knowledge, this is the first exact result for the optical conductivity of a realistic strongly interacting QCP, and will thus
serve as a benchmark for the dynamical response of quantum critical systems. We note that  
(\ref{SigmaInfWZ}) is both larger than the Dirac fermion conductivity $\sigma_\infty= \frac{1}{16} \!=\! 0.0625$~\cite{osborn1994} and the conductivity 
of the Cooper pair superconductor-insulator QCP $\sigma_\infty=0.226$~\cite{gazit2013,chen2014,witczak-krempa2014,katz2014,gazit2014,kos2015}. 
Our result (\ref{SigmaInfWZ}) is tantalizingly close to the latter, suggesting 
that even though the Dirac semimetal-superconductor QCP naively seems to have more conducting degrees of freedom, these interact more strongly. 
To put our exact result in perspective, we emphasize that $\sigma_\infty$ for the superconductor-insulator QCP has been the subject of numerous
studies~\cite{fisher1990,cha1991,fazio1996,cha1991,smakov2005} over the past three decades
but was reliably obtained only recently via large-scale quantum Monte Carlo simulations~\cite{gazit2013,chen2014,witczak-krempa2014,katz2014,gazit2014} 
and the conformal bootstrap approach~\cite{kos2015}. 
Finally, note that (\ref{SigmaInfWZ}) is smaller than the conductivity of the Gaussian  
fixed point, $\tfrac{5}{16} \!=\! 0.3125$, in agreement with the expectation that strong interactions reduce the charge mobility.             


{\bf Entanglement entropy}:      
There is currently much interest in the entanglement properties of QCPs~\cite{vidal2003,calabrese2004}. In particular,  
the ground state entanglement entropy across a spatial region containing a sharp corner with opening angle $\theta$ contains a subleading 
logarithmic term whose coefficient $a(\theta)$ depends only on the universality class of the QCP. This coefficient constitutes a new 
measure of the gapless degrees of freedom in strongly interacting systems.
Recent numerical work has focused on determining $a(\theta)$ for various interacting 2D QCPs, such as the XY and
Heisenberg QCPs appearing in theories of quantum magnetism~\cite{stoudenmire2014,kallin2014,wessel15}.  
For QCPs with emergent Lorentz invariance, the behavior of $a(\theta)$ near $\theta\!=\!\pi$ is  
determined by the stress-tensor correlation coefficient $C_T$ encountered above~\cite{bueno2015,faulkner15}, 
\begin{align} \label{atheta}
a(\theta)\approx\lambda_{\rm corner} (\pi-\theta)^2,\hspace{5mm}  
\lambda_{\rm corner}=\pi^2 C_T/24.
\end{align}  
Using our exact result for $\sigma_\infty$, we obtain an exact result in closed form for the corner coefficient of the 
semimetal-superconductor QCP occurring on the surface of a topological insulator:
$\lambda_{\rm corner}=\sigma_\infty/20=\frac{16\pi-9\sqrt{3}}{972\pi}\!\approx\! 0.01136$. Unexpectedly, the optical conductivity at zero temperature entirely determines this property of the entanglement entropy. These two quantities are
generally unrelated in the absence of supersymmetry, as can be seen in Table~\ref{tbl:ratio}. 
We note that an integral expression for $\lambda_{\rm corner}$ has been given previously~\cite{bueno1505}.      
In addition, our result for $\lambda_{\rm corner}$ leads to an exact lower bound on $a(\theta)$ for \emph{all} 
opening angles~\cite{bound}: $a(\theta)\geq (2\sigma_\infty/5)\ln[1/\sin(\theta/2)]$.

{\bf Optical conductivity at finite temperature}:   
So far our discussion has centered on $T=0$ properties. We now study the 
finite-$T$ optical conductivity. The most reliable statements can be made
in the regime $k_B T\ll \hbar\omega$ corresponding   
to the response at temperatures much lower than the measurement frequency, where one obtains the nontrivial expansion~\cite{katz2014} 
\begin{align}\label{conduc}
\frac{\sigma(\omega,T)}{e^2/\hbar} = \sigma_\infty+b\left(\frac{ik_B T}{\hbar\omega}\right)^{\!\!3-1/\nu} \!\!\!   
  +b'\left(\frac{ik_B T}{\hbar \omega}\right)^{\!3} \!+\dotsb 
\end{align}  
where the dots denote higher powers of $k_BT/\hbar\omega$, corresponding to increasingly small corrections. The dimensionless real coefficients $b,b'$  
are universal properties of the QCP, and $\nu$ is the correlation-length critical exponent.   
The structure of (\ref{conduc}) follows from simple physical arguments, which we now briefly review. 
The large frequency expansion follows from the short time expansion of the operator product $J_x(t)J_x(0)$  
appearing in the Kubo formula for the conductivity. As $t\to 0$, one can replace the product by a series involving operators
of increasing scaling dimensions~\cite{katz2014}, called the operator product expansion (OPE). 
The operators that dominate the expansion are the identity, the ``mass'' operator $|\phi|^2$ that tunes the system to the QCP in the Landau-Ginzburg Lagrangian (\ref{L}), and the stress tensor $T_{\mu\nu}$. We can thus schematically write
$J_xJ_x\sim  1+|\phi|^2+T_{\mu\nu}+\dotsb$. The coefficients that multiply each operator in the series, omitted in this schematic expansion, are called OPE coefficients.
The parameters $b,b'$ are proportional to the OPE coefficients multiplying $|\phi|^2$ and $T_{\mu\nu}$, respectively.  
The corresponding powers of $k_BT/\hbar\omega$ in \req{conduc} are the scaling dimensions of these operators. 
The dimension of $|\phi|^2$ is $\Delta_r\!=\!3-1/\nu$, where the correlation length exponent $\nu$ can be estimated
via the $\epsilon$ expansion, $\nu \approx 0.75$~\cite{ScottThomas}.   
A more accurate result is given by the conformal bootstrap, which predicts $\Delta_r=1.9098(20)$~\cite{bobev2015}. 
In contrast, the stress tensor is conserved
and its scaling dimension is not renormalized: $\Delta_T=3$.  

Turning to the coefficients in Eq.~(\ref{conduc}), SUSY does not impose any constraints on $b$.  
However, in the case of $b'$ SUSY leads to the strong result:  
\begin{align} 
  b' = 0\,.  
\end{align} 
To understand this result, recall that $b'\propto\gamma$, where $\gamma$ is an OPE coefficient multiplying the stress tensor. This latter coefficient can be determined from the three-point correlation function $\langle T_{\mu\nu}J_\lambda J_\rho\rangle$ at zero temperature~\cite{katz2014}. To see if SUSY constrains $\gamma$, we use a recent result for the general form of the three-point correlation function $\langle\mathcal{J}_\nu \mathcal{J}_\lambda \mathcal{J}_\rho\rangle$ of the supercurrent~\cite{buchbinder2015}. While  
the precise form of this function is fairly complicated, its crucial feature is that it is characterized by a single overall  
constant, analogously to the two-point correlation function of the supercurrent. By extracting the $\langle TJJ\rangle$ component of the three-point correlation function of the supercurrent, 
we find that $\gamma$ and thus $b'$ vanish identically~\cite{SuppMat}. As shown in Table~\ref{tbl:ratio}, this
is not the case at the superconductor-insulator QCP of Cooper pairs~\cite{katz2014}, as expected in the absence of emergent SUSY.  

{\bf Sum rules}: From the point of view of the frequency dependence, the finite-temperature results we have given so far for the optical conductivity correspond to the high-frequency regime $\hbar\omega\gg k_BT$. In fact, we have sufficient information about the QCP to go even further and constrain the integral of the finite-temperature optical conductivity over \emph{all} frequencies by way of a sum rule~\cite{herzog2011,will_qnm,katz2014}:
\begin{align} \label{sr}
  \int_0^\infty d\omega\, \left(\text{Re}\,\sigma(\omega,T) - \sigma_\infty e^2/\hbar \right) = 0\,.    
\end{align}
A dual sum rule obtained by replacing $\sigma$ with $1/\sigma$ also holds~\cite{will_qnm,will_qnm}. 
The key point is that the integrand must decay sufficiently fast at high frequencies.  
This is the case here, since in that limit the integrand scales as $(T/\omega)^{3-1/\nu}$ [Eq.~(\ref{conduc})], and 
we know that $\nu>1/2$~\cite{bobev2015}.        

{\bf Experimental realizations}: 
Recent experiments suggest that intrinsic (as opposed to proximity-induced) superconductivity may have been observed on the surface of the 3D topological insulator Sb$_2$Te$_3$~\cite{zhao2015}. Scanning tunneling microscopy data suggests an inhomogeneous distribution of local critical temperatures $T_c(\b{r})$ as high as 60~K, with global phase coherence achieved only at a much lower $\sim 9$~K. The QCP discussed here remains stable against quenched disorder in $T_c$, assuming it is short-ranged, only if the Harris criterion $\nu d>2$ is satisfied, where $d\!=\!2$ is the spatial dimension and $\nu$ is the correlation length exponent of the clean QCP~\cite{harris1974}. Using the conformal bootstrap result quoted earlier, one obtains $\nu\approx 0.917$, implying that the QCP is compromised by this type of disorder. Signatures of the clean QCP will nevertheless be observable above the crossover temperature $k_B T^*\sim\Lambda W^{1/(2/\nu-d)}\sim\Lambda W^{5.5}$ where $\Lambda$ is a high-energy cutoff that can be taken as the bulk gap of the topological insulator and $W$ is some dimensionless measure of the disorder strength~\cite{nandkishore2013}. Given the high power of $W$, 
one expects that the $\hbar\omega\gg k_B T> k_B T^*$ regime---in which the results discussed here hold---will be reachable in the near future in 
samples with moderate amounts of disorder.  

{\bf Discussion \& outlook}:  
We have analyzed the dynamical response properties of a strongly interacting QCP occurring on the surface of a 3D topological insulator between the gapless Dirac surface state and a gapped surface superconductor. The emergence of SUSY in the low-energy limit at this QCP allowed us to deduce exact results for the dynamical response of the system in closed form, as summarized in Table~\ref{tbl:ratio}. We found that the zero-temperature optical conductivity and dynamical shear viscosity coefficient are frequency-independent, proportional to each other, and given by a simple irrational number, Eq.~(\ref{SigmaInfWZ}). We further made exact statements concerning the finite-temperature optical conductivity, including high-frequency asymptotics and sum rules. It is natural to ask if other properties of this QCP can be deduced from SUSY, such as the entanglement R\'enyi entropies of corners~\cite{twist}. More broadly, it would be worthwhile to investigate other QCPs with emergent SUSY in both two and three spatial dimensions.  
\\ 
\begin{acknowledgments} 
\emph{Acknowledgements} ---
We are grateful to D. Maz\'{a}\v{c} and S. Sachdev for insightful observations.
We would also like to thank I. Affleck, V. Bouchard, S. Gopalakrishnan, L. LeBlanc, T. Nishioka, A. Penin, C. Quigley, X. Yin, and N. Zerf for useful discussions. 
W.W.-K. was supported in part by a postdoctoral fellowship from NSERC. 
J.M. was supported by NSERC grant \#RGPIN-2014-4608, the CRC Program, CIFAR, and the University of Alberta, and wishes to acknowledge the hospitality of the Perimeter Institute for Theoretical Physics (PI) and the Kavli Institute for Theoretical Physics (KITP) where part of this research was carried out. Research at PI is supported by the Government of Canada through Industry Canada and by the Province of Ontario through the Ministry of Economic Development \& Innovation. This research was also supported in part by the National Science Foundation under Grant No.~NSF PHY11-25915 (KITP).
\end{acknowledgments} 

\bibliography{susy}

\begin{thebibliography}{54}%
\makeatletter
\providecommand \@ifxundefined [1]{%
 \@ifx{#1\undefined}
}%
\providecommand \@ifnum [1]{%
 \ifnum #1\expandafter \@firstoftwo
 \else \expandafter \@secondoftwo
 \fi
}%
\providecommand \@ifx [1]{%
 \ifx #1\expandafter \@firstoftwo
 \else \expandafter \@secondoftwo
 \fi
}%
\providecommand \natexlab [1]{#1}%
\providecommand \enquote  [1]{``#1''}%
\providecommand \bibnamefont  [1]{#1}%
\providecommand \bibfnamefont [1]{#1}%
\providecommand \citenamefont [1]{#1}%
\providecommand \href@noop [0]{\@secondoftwo}%
\providecommand \href [0]{\begingroup \@sanitize@url \@href}%
\providecommand \@href[1]{\@@startlink{#1}\@@href}%
\providecommand \@@href[1]{\endgroup#1\@@endlink}%
\providecommand \@sanitize@url [0]{\catcode `\\12\catcode `\$12\catcode
  `\&12\catcode `\#12\catcode `\^12\catcode `\_12\catcode `\%12\relax}%
\providecommand \@@startlink[1]{}%
\providecommand \@@endlink[0]{}%
\providecommand \url  [0]{\begingroup\@sanitize@url \@url }%
\providecommand \@url [1]{\endgroup\@href {#1}{\urlprefix }}%
\providecommand \urlprefix  [0]{URL }%
\providecommand \Eprint [0]{\href }%
\providecommand \doibase [0]{http://dx.doi.org/}%
\providecommand \selectlanguage [0]{\@gobble}%
\providecommand \bibinfo  [0]{\@secondoftwo}%
\providecommand \bibfield  [0]{\@secondoftwo}%
\providecommand \translation [1]{[#1]}%
\providecommand \BibitemOpen [0]{}%
\providecommand \bibitemStop [0]{}%
\providecommand \bibitemNoStop [0]{.\EOS\space}%
\providecommand \EOS [0]{\spacefactor3000\relax}%
\providecommand \BibitemShut  [1]{\csname bibitem#1\endcsname}%
\let\auto@bib@innerbib\@empty
\bibitem [{\citenamefont {Hasan}\ and\ \citenamefont {Kane}(2010)}]{RMP_TI}%
  \BibitemOpen
  \bibfield  {author} {\bibinfo {author} {\bibfnamefont {M.~Z.}\ \bibnamefont
  {Hasan}}\ and\ \bibinfo {author} {\bibfnamefont {C.~L.}\ \bibnamefont
  {Kane}},\ }\href {\doibase 10.1103/RevModPhys.82.3045} {\bibfield  {journal}
  {\bibinfo  {journal} {Rev. Mod. Phys.}\ }\textbf {\bibinfo {volume} {82}},\
  \bibinfo {pages} {3045} (\bibinfo {year} {2010})}\BibitemShut {NoStop}%
\bibitem [{\citenamefont {{Qi}}\ and\ \citenamefont {{Zhang}}(2011)}]{RMP_TI2}%
  \BibitemOpen
  \bibfield  {author} {\bibinfo {author} {\bibfnamefont {X.-L.}\ \bibnamefont
  {{Qi}}}\ and\ \bibinfo {author} {\bibfnamefont {S.-C.}\ \bibnamefont
  {{Zhang}}},\ }\href {\doibase 10.1103/RevModPhys.83.1057} {\bibfield
  {journal} {\bibinfo  {journal} {Rev. Mod. Phys.}\ }\textbf {\bibinfo {volume}
  {83}},\ \bibinfo {pages} {1057} (\bibinfo {year} {2011})}\BibitemShut
  {NoStop}%
\bibitem [{\citenamefont {{Xu}}(2010)}]{cenke1}%
  \BibitemOpen
  \bibfield  {author} {\bibinfo {author} {\bibfnamefont {C.}~\bibnamefont
  {{Xu}}},\ }\href {\doibase 10.1103/PhysRevB.81.054403} {\bibfield  {journal}
  {\bibinfo  {journal} {Phys. Rev. B}\ }\textbf {\bibinfo {volume} {81}},\
  \bibinfo {eid} {054403} (\bibinfo {year} {2010})}\BibitemShut {NoStop}%
\bibitem [{\citenamefont {Xu}(2010)}]{cenke2}%
  \BibitemOpen
  \bibfield  {author} {\bibinfo {author} {\bibfnamefont {C.}~\bibnamefont
  {Xu}},\ }\href {\doibase 10.1103/PhysRevB.81.020411} {\bibfield  {journal}
  {\bibinfo  {journal} {Phys. Rev. B}\ }\textbf {\bibinfo {volume} {81}},\
  \bibinfo {pages} {020411} (\bibinfo {year} {2010})}\BibitemShut {NoStop}%
\bibitem [{\citenamefont {Lundgren}\ and\ \citenamefont
  {Maciejko}(2015)}]{lundgren}%
  \BibitemOpen
  \bibfield  {author} {\bibinfo {author} {\bibfnamefont {R.}~\bibnamefont
  {Lundgren}}\ and\ \bibinfo {author} {\bibfnamefont {J.}~\bibnamefont
  {Maciejko}},\ }\href {\doibase 10.1103/PhysRevLett.115.066401} {\bibfield
  {journal} {\bibinfo  {journal} {Phys. Rev. Lett.}\ }\textbf {\bibinfo
  {volume} {115}},\ \bibinfo {eid} {066401} (\bibinfo {year}
  {2015})}\BibitemShut {NoStop}%
\bibitem [{\citenamefont {Grover}\ \emph {et~al.}(2014)\citenamefont {Grover},
  \citenamefont {Sheng},\ and\ \citenamefont {Vishwanath}}]{grover2014}%
  \BibitemOpen
  \bibfield  {author} {\bibinfo {author} {\bibfnamefont {T.}~\bibnamefont
  {Grover}}, \bibinfo {author} {\bibfnamefont {D.~N.}\ \bibnamefont {Sheng}}, \
  and\ \bibinfo {author} {\bibfnamefont {A.}~\bibnamefont {Vishwanath}},\
  }\href {\doibase 10.1126/science.1248253} {\bibfield  {journal} {\bibinfo
  {journal} {Science}\ }\textbf {\bibinfo {volume} {344}},\ \bibinfo {pages}
  {280} (\bibinfo {year} {2014})}\BibitemShut {NoStop}%
\bibitem [{\citenamefont {Ponte}\ and\ \citenamefont {Lee}(2014)}]{ponte2014}%
  \BibitemOpen
  \bibfield  {author} {\bibinfo {author} {\bibfnamefont {P.}~\bibnamefont
  {Ponte}}\ and\ \bibinfo {author} {\bibfnamefont {S.-S.}\ \bibnamefont
  {Lee}},\ }\href {\doibase 10.1088/1367-2630/16/1/013044} {\bibfield
  {journal} {\bibinfo  {journal} {New J. Phys.}\ }\textbf {\bibinfo {volume}
  {16}},\ \bibinfo {pages} {013044} (\bibinfo {year} {2014})}\BibitemShut
  {NoStop}%
\bibitem [{\citenamefont {Zhao}\ \emph {et~al.}(2015)\citenamefont {Zhao},
  \citenamefont {Deng}, \citenamefont {Korzhovska}, \citenamefont
  {Begliarbekov}, \citenamefont {Chen}, \citenamefont {Andrade}, \citenamefont
  {Rosenthal}, \citenamefont {Pasupathy}, \citenamefont {Oganesyan},\ and\
  \citenamefont {Krusin-Elbaum}}]{zhao2015}%
  \BibitemOpen
  \bibfield  {author} {\bibinfo {author} {\bibfnamefont {L.}~\bibnamefont
  {Zhao}}, \bibinfo {author} {\bibfnamefont {H.}~\bibnamefont {Deng}}, \bibinfo
  {author} {\bibfnamefont {I.}~\bibnamefont {Korzhovska}}, \bibinfo {author}
  {\bibfnamefont {M.}~\bibnamefont {Begliarbekov}}, \bibinfo {author}
  {\bibfnamefont {Z.}~\bibnamefont {Chen}}, \bibinfo {author} {\bibfnamefont
  {E.}~\bibnamefont {Andrade}}, \bibinfo {author} {\bibfnamefont
  {E.}~\bibnamefont {Rosenthal}}, \bibinfo {author} {\bibfnamefont
  {A.}~\bibnamefont {Pasupathy}}, \bibinfo {author} {\bibfnamefont
  {V.}~\bibnamefont {Oganesyan}}, \ and\ \bibinfo {author} {\bibfnamefont
  {L.}~\bibnamefont {Krusin-Elbaum}},\ }\href {\doibase 10.1038/ncomms9279}
  {\bibfield  {journal} {\bibinfo  {journal} {Nature Commun.}\ }\textbf
  {\bibinfo {volume} {6}},\ \bibinfo {pages} {8279} (\bibinfo {year}
  {2015})}\BibitemShut {NoStop}%
\bibitem [{\citenamefont {Lee}(2007)}]{lee2007}%
  \BibitemOpen
  \bibfield  {author} {\bibinfo {author} {\bibfnamefont {S.-S.}\ \bibnamefont
  {Lee}},\ }\href {\doibase 10.1103/PhysRevB.76.075103} {\bibfield  {journal}
  {\bibinfo  {journal} {Phys. Rev. B}\ }\textbf {\bibinfo {volume} {76}},\
  \bibinfo {pages} {075103} (\bibinfo {year} {2007})}\BibitemShut {NoStop}%
\bibitem [{\citenamefont {Jian}\ \emph {et~al.}(2015)\citenamefont {Jian},
  \citenamefont {Jiang},\ and\ \citenamefont {Yao}}]{jian2015}%
  \BibitemOpen
  \bibfield  {author} {\bibinfo {author} {\bibfnamefont {S.-K.}\ \bibnamefont
  {Jian}}, \bibinfo {author} {\bibfnamefont {Y.-F.}\ \bibnamefont {Jiang}}, \
  and\ \bibinfo {author} {\bibfnamefont {H.}~\bibnamefont {Yao}},\ }\href
  {\doibase 10.1103/PhysRevLett.114.237001} {\bibfield  {journal} {\bibinfo
  {journal} {Phys. Rev. Lett.}\ }\textbf {\bibinfo {volume} {114}},\ \bibinfo
  {pages} {237001} (\bibinfo {year} {2015})}\BibitemShut {NoStop}%
\bibitem [{\citenamefont {Yu}\ and\ \citenamefont {Yang}(2010)}]{yu2010}%
  \BibitemOpen
  \bibfield  {author} {\bibinfo {author} {\bibfnamefont {Y.}~\bibnamefont
  {Yu}}\ and\ \bibinfo {author} {\bibfnamefont {K.}~\bibnamefont {Yang}},\
  }\href {\doibase 10.1103/PhysRevLett.105.150605} {\bibfield  {journal}
  {\bibinfo  {journal} {Phys. Rev. Lett.}\ }\textbf {\bibinfo {volume} {105}},\
  \bibinfo {pages} {150605} (\bibinfo {year} {2010})}\BibitemShut {NoStop}%
\bibitem [{Sco()}]{ScottThomas}%
  \BibitemOpen
  \href@noop {} {}\bibinfo {note} {S. Thomas, talk at the 2005 KITP Conference
  on Quantum Phase Transitions, Kavli Institute for Theoretical Physics, Santa
  Barbara, 21 January 2005.}\BibitemShut {Stop}%
\bibitem [{\citenamefont {Aharony}\ \emph {et~al.}(1997)\citenamefont
  {Aharony}, \citenamefont {Hanany}, \citenamefont {Intriligator},
  \citenamefont {Seiberg},\ and\ \citenamefont {Strassler}}]{aharony1997}%
  \BibitemOpen
  \bibfield  {author} {\bibinfo {author} {\bibfnamefont {O.}~\bibnamefont
  {Aharony}}, \bibinfo {author} {\bibfnamefont {A.}~\bibnamefont {Hanany}},
  \bibinfo {author} {\bibfnamefont {K.}~\bibnamefont {Intriligator}}, \bibinfo
  {author} {\bibfnamefont {N.}~\bibnamefont {Seiberg}}, \ and\ \bibinfo
  {author} {\bibfnamefont {M.~J.}\ \bibnamefont {Strassler}},\ }\href {\doibase
  10.1016/S0550-3213(97)00323-4} {\bibfield  {journal} {\bibinfo  {journal}
  {Nucl. Phys. B}\ }\textbf {\bibinfo {volume} {499}},\ \bibinfo {pages} {67}
  (\bibinfo {year} {1997})}\BibitemShut {NoStop}%
\bibitem [{\citenamefont {Fisher}\ \emph {et~al.}(1990)\citenamefont {Fisher},
  \citenamefont {Grinstein},\ and\ \citenamefont {Girvin}}]{fisher1990}%
  \BibitemOpen
  \bibfield  {author} {\bibinfo {author} {\bibfnamefont {M.~P.~A.}\
  \bibnamefont {Fisher}}, \bibinfo {author} {\bibfnamefont {G.}~\bibnamefont
  {Grinstein}}, \ and\ \bibinfo {author} {\bibfnamefont {S.~M.}\ \bibnamefont
  {Girvin}},\ }\href {\doibase 10.1103/PhysRevLett.64.587} {\bibfield
  {journal} {\bibinfo  {journal} {Phys. Rev. Lett.}\ }\textbf {\bibinfo
  {volume} {64}},\ \bibinfo {pages} {587} (\bibinfo {year} {1990})}\BibitemShut
  {NoStop}%
\bibitem [{\citenamefont {Damle}\ and\ \citenamefont
  {Sachdev}(1997)}]{damle1997}%
  \BibitemOpen
  \bibfield  {author} {\bibinfo {author} {\bibfnamefont {K.}~\bibnamefont
  {Damle}}\ and\ \bibinfo {author} {\bibfnamefont {S.}~\bibnamefont
  {Sachdev}},\ }\href {\doibase 10.1103/PhysRevB.56.8714} {\bibfield  {journal}
  {\bibinfo  {journal} {Phys. Rev. B}\ }\textbf {\bibinfo {volume} {56}},\
  \bibinfo {pages} {8714} (\bibinfo {year} {1997})}\BibitemShut {NoStop}%
\bibitem [{\citenamefont {Closset}\ \emph {et~al.}(2013)\citenamefont
  {Closset}, \citenamefont {Dumitrescu}, \citenamefont {Festuccia},\ and\
  \citenamefont {Komargodski}}]{closset2013}%
  \BibitemOpen
  \bibfield  {author} {\bibinfo {author} {\bibfnamefont {C.}~\bibnamefont
  {Closset}}, \bibinfo {author} {\bibfnamefont {T.~T.}\ \bibnamefont
  {Dumitrescu}}, \bibinfo {author} {\bibfnamefont {G.}~\bibnamefont
  {Festuccia}}, \ and\ \bibinfo {author} {\bibfnamefont {Z.}~\bibnamefont
  {Komargodski}},\ }\href {\doibase 10.1007/JHEP05(2013)017} {\bibfield
  {journal} {\bibinfo  {journal} {JHEP}\ }\textbf {\bibinfo {volume} {1305}},\
  \bibinfo {pages} {017} (\bibinfo {year} {2013})}\BibitemShut {NoStop}%
\bibitem [{\citenamefont {Osborn}(1999)}]{osborn1999}%
  \BibitemOpen
  \bibfield  {author} {\bibinfo {author} {\bibfnamefont {H.}~\bibnamefont
  {Osborn}},\ }\href {\doibase 10.1006/aphy.1998.5893} {\bibfield  {journal}
  {\bibinfo  {journal} {Ann. Phys.}\ }\textbf {\bibinfo {volume} {272}},\
  \bibinfo {pages} {243} (\bibinfo {year} {1999})}\BibitemShut {NoStop}%
\bibitem [{\citenamefont {Barnes}\ \emph {et~al.}(2005)\citenamefont {Barnes},
  \citenamefont {Gorbatov}, \citenamefont {Intriligator}, \citenamefont
  {Sudano},\ and\ \citenamefont {Wright}}]{barnes2005}%
  \BibitemOpen
  \bibfield  {author} {\bibinfo {author} {\bibfnamefont {E.}~\bibnamefont
  {Barnes}}, \bibinfo {author} {\bibfnamefont {E.}~\bibnamefont {Gorbatov}},
  \bibinfo {author} {\bibfnamefont {K.}~\bibnamefont {Intriligator}}, \bibinfo
  {author} {\bibfnamefont {M.}~\bibnamefont {Sudano}}, \ and\ \bibinfo {author}
  {\bibfnamefont {J.}~\bibnamefont {Wright}},\ }\href {\doibase
  10.1016/j.nuclphysb.2005.10.003} {\bibfield  {journal} {\bibinfo  {journal}
  {Nucl. Phys. B}\ }\textbf {\bibinfo {volume} {730}},\ \bibinfo {pages} {210}
  (\bibinfo {year} {2005})}\BibitemShut {NoStop}%
\bibitem [{\citenamefont {Buchbinder}\ \emph {et~al.}(2015)\citenamefont
  {Buchbinder}, \citenamefont {Kuzenko},\ and\ \citenamefont
  {Samsonov}}]{buchbinder2015}%
  \BibitemOpen
  \bibfield  {author} {\bibinfo {author} {\bibfnamefont {E.~I.}\ \bibnamefont
  {Buchbinder}}, \bibinfo {author} {\bibfnamefont {S.~M.}\ \bibnamefont
  {Kuzenko}}, \ and\ \bibinfo {author} {\bibfnamefont {I.~B.}\ \bibnamefont
  {Samsonov}},\ }\href {\doibase 10.1007/JHEP06(2015)138} {\bibfield  {journal}
  {\bibinfo  {journal} {JHEP}\ }\textbf {\bibinfo {volume} {1506}},\ \bibinfo
  {pages} {138} (\bibinfo {year} {2015})}\BibitemShut {NoStop}%
\bibitem [{\citenamefont {Osborn}\ and\ \citenamefont
  {Petkou}(1994)}]{osborn1994}%
  \BibitemOpen
  \bibfield  {author} {\bibinfo {author} {\bibfnamefont {H.}~\bibnamefont
  {Osborn}}\ and\ \bibinfo {author} {\bibfnamefont {A.}~\bibnamefont
  {Petkou}},\ }\href {\doibase 10.1006/aphy.1994.1045} {\bibfield  {journal}
  {\bibinfo  {journal} {Ann. Phys.}\ }\textbf {\bibinfo {volume} {231}},\
  \bibinfo {pages} {311} (\bibinfo {year} {1994})}\BibitemShut {NoStop}%
\bibitem [{Sup()}]{SuppMat}%
  \BibitemOpen
  \href@noop {} {}\bibinfo {note} {See Supplemental Material, which includes
  Refs.~\cite{dumitrescu2011, hofman08, myers2011, chowdhury2013}}\BibitemShut
  {NoStop}%
\bibitem [{\citenamefont {Dumitrescu}\ and\ \citenamefont
  {Seiberg}(2011)}]{dumitrescu2011}%
  \BibitemOpen
  \bibfield  {author} {\bibinfo {author} {\bibfnamefont {T.~T.}\ \bibnamefont
  {Dumitrescu}}\ and\ \bibinfo {author} {\bibfnamefont {N.}~\bibnamefont
  {Seiberg}},\ }\href {\doibase 10.1007/JHEP07(2011)095} {\bibfield  {journal}
  {\bibinfo  {journal} {JHEP}\ }\textbf {\bibinfo {volume} {1107}},\ \bibinfo
  {pages} {095} (\bibinfo {year} {2011})}\BibitemShut {NoStop}%
\bibitem [{\citenamefont {{Hofman}}\ and\ \citenamefont
  {{Maldacena}}(2008)}]{hofman08}%
  \BibitemOpen
  \bibfield  {author} {\bibinfo {author} {\bibfnamefont {D.~M.}\ \bibnamefont
  {{Hofman}}}\ and\ \bibinfo {author} {\bibfnamefont {J.}~\bibnamefont
  {{Maldacena}}},\ }\href {\doibase 10.1088/1126-6708/2008/05/012} {\bibfield
  {journal} {\bibinfo  {journal} {JHEP}\ }\textbf {\bibinfo {volume} {0805}},\
  \bibinfo {eid} {012} (\bibinfo {year} {2008})}\BibitemShut {NoStop}%
\bibitem [{\citenamefont {{Myers}}\ \emph {et~al.}(2011)\citenamefont
  {{Myers}}, \citenamefont {{Sachdev}},\ and\ \citenamefont
  {{Singh}}}]{myers2011}%
  \BibitemOpen
  \bibfield  {author} {\bibinfo {author} {\bibfnamefont {R.~C.}\ \bibnamefont
  {{Myers}}}, \bibinfo {author} {\bibfnamefont {S.}~\bibnamefont {{Sachdev}}},
  \ and\ \bibinfo {author} {\bibfnamefont {A.}~\bibnamefont {{Singh}}},\ }\href
  {\doibase 10.1103/PhysRevD.83.066017} {\bibfield  {journal} {\bibinfo
  {journal} {\prd}\ }\textbf {\bibinfo {volume} {83}},\ \bibinfo {eid} {066017}
  (\bibinfo {year} {2011})}\BibitemShut {NoStop}%
\bibitem [{\citenamefont {Chowdhury}\ \emph {et~al.}(2013)\citenamefont
  {Chowdhury}, \citenamefont {Raju}, \citenamefont {Sachdev}, \citenamefont
  {Singh},\ and\ \citenamefont {Strack}}]{chowdhury2013}%
  \BibitemOpen
  \bibfield  {author} {\bibinfo {author} {\bibfnamefont {D.}~\bibnamefont
  {Chowdhury}}, \bibinfo {author} {\bibfnamefont {S.}~\bibnamefont {Raju}},
  \bibinfo {author} {\bibfnamefont {S.}~\bibnamefont {Sachdev}}, \bibinfo
  {author} {\bibfnamefont {A.}~\bibnamefont {Singh}}, \ and\ \bibinfo {author}
  {\bibfnamefont {P.}~\bibnamefont {Strack}},\ }\href {\doibase
  10.1103/PhysRevB.87.085138} {\bibfield  {journal} {\bibinfo  {journal} {Phy.
  Rev. B}\ }\textbf {\bibinfo {volume} {87}},\ \bibinfo {pages} {085138}
  (\bibinfo {year} {2013})}\BibitemShut {NoStop}%
\bibitem [{\citenamefont {Bradlyn}\ \emph {et~al.}(2012)\citenamefont
  {Bradlyn}, \citenamefont {Goldstein},\ and\ \citenamefont
  {Read}}]{bradlyn2012}%
  \BibitemOpen
  \bibfield  {author} {\bibinfo {author} {\bibfnamefont {B.}~\bibnamefont
  {Bradlyn}}, \bibinfo {author} {\bibfnamefont {M.}~\bibnamefont {Goldstein}},
  \ and\ \bibinfo {author} {\bibfnamefont {N.}~\bibnamefont {Read}},\ }\href
  {\doibase 10.1103/PhysRevB.86.245309} {\bibfield  {journal} {\bibinfo
  {journal} {Phys. Rev. B}\ }\textbf {\bibinfo {volume} {86}},\ \bibinfo
  {pages} {245309} (\bibinfo {year} {2012})}\BibitemShut {NoStop}%
\bibitem [{\citenamefont {Witczak-Krempa}(2015)}]{witczak-krempa2015}%
  \BibitemOpen
  \bibfield  {author} {\bibinfo {author} {\bibfnamefont {W.}~\bibnamefont
  {Witczak-Krempa}},\ }\href {\doibase 10.1103/PhysRevLett.114.177201}
  {\bibfield  {journal} {\bibinfo  {journal} {Phys. Rev. Lett.}\ }\textbf
  {\bibinfo {volume} {114}},\ \bibinfo {pages} {177201} (\bibinfo {year}
  {2015})}\BibitemShut {NoStop}%
\bibitem [{\citenamefont {Imamura}(2011)}]{imamura2011}%
  \BibitemOpen
  \bibfield  {author} {\bibinfo {author} {\bibfnamefont {Y.}~\bibnamefont
  {Imamura}},\ }\href {\doibase 10.1007/JHEP09(2011)133} {\bibfield  {journal}
  {\bibinfo  {journal} {JHEP}\ }\textbf {\bibinfo {volume} {1109}},\ \bibinfo
  {pages} {133} (\bibinfo {year} {2011})}\BibitemShut {NoStop}%
\bibitem [{\citenamefont {Imamura}\ and\ \citenamefont
  {Yokoyama}(2012)}]{imamura2012}%
  \BibitemOpen
  \bibfield  {author} {\bibinfo {author} {\bibfnamefont {Y.}~\bibnamefont
  {Imamura}}\ and\ \bibinfo {author} {\bibfnamefont {D.}~\bibnamefont
  {Yokoyama}},\ }\href {\doibase 10.1103/PhysRevD.85.025015} {\bibfield
  {journal} {\bibinfo  {journal} {Phys. Rev. D}\ }\textbf {\bibinfo {volume}
  {85}},\ \bibinfo {pages} {025015} (\bibinfo {year} {2012})}\BibitemShut
  {NoStop}%
\bibitem [{\citenamefont {Nishioka}\ and\ \citenamefont
  {Yonekura}(2013)}]{nishioka2013}%
  \BibitemOpen
  \bibfield  {author} {\bibinfo {author} {\bibfnamefont {T.}~\bibnamefont
  {Nishioka}}\ and\ \bibinfo {author} {\bibfnamefont {K.}~\bibnamefont
  {Yonekura}},\ }\href {\doibase 10.1007/JHEP05(2013)165} {\bibfield  {journal}
  {\bibinfo  {journal} {JHEP}\ }\textbf {\bibinfo {volume} {1305}},\ \bibinfo
  {pages} {165} (\bibinfo {year} {2013})}\BibitemShut {NoStop}%
\bibitem [{\citenamefont {Gazit}\ \emph {et~al.}(2013)\citenamefont {Gazit},
  \citenamefont {Podolsky}, \citenamefont {Auerbach},\ and\ \citenamefont
  {Arovas}}]{gazit2013}%
  \BibitemOpen
  \bibfield  {author} {\bibinfo {author} {\bibfnamefont {S.}~\bibnamefont
  {Gazit}}, \bibinfo {author} {\bibfnamefont {D.}~\bibnamefont {Podolsky}},
  \bibinfo {author} {\bibfnamefont {A.}~\bibnamefont {Auerbach}}, \ and\
  \bibinfo {author} {\bibfnamefont {D.~P.}\ \bibnamefont {Arovas}},\ }\href
  {\doibase 10.1103/PhysRevB.88.235108} {\bibfield  {journal} {\bibinfo
  {journal} {Phys. Rev. B}\ }\textbf {\bibinfo {volume} {88}},\ \bibinfo
  {pages} {235108} (\bibinfo {year} {2013})}\BibitemShut {NoStop}%
\bibitem [{\citenamefont {Chen}\ \emph {et~al.}(2014)\citenamefont {Chen},
  \citenamefont {Liu}, \citenamefont {Deng}, \citenamefont {Pollet},\ and\
  \citenamefont {{Prokof'ev}}}]{chen2014}%
  \BibitemOpen
  \bibfield  {author} {\bibinfo {author} {\bibfnamefont {K.}~\bibnamefont
  {Chen}}, \bibinfo {author} {\bibfnamefont {L.}~\bibnamefont {Liu}}, \bibinfo
  {author} {\bibfnamefont {Y.}~\bibnamefont {Deng}}, \bibinfo {author}
  {\bibfnamefont {L.}~\bibnamefont {Pollet}}, \ and\ \bibinfo {author}
  {\bibfnamefont {N.}~\bibnamefont {{Prokof'ev}}},\ }\href {\doibase
  10.1103/PhysRevLett.112.030402} {\bibfield  {journal} {\bibinfo  {journal}
  {Phys. Rev. Lett.}\ }\textbf {\bibinfo {volume} {112}},\ \bibinfo {pages}
  {030402} (\bibinfo {year} {2014})}\BibitemShut {NoStop}%
\bibitem [{\citenamefont {Witczak-Krempa}\ \emph {et~al.}(2014)\citenamefont
  {Witczak-Krempa}, \citenamefont {{S{\o}rensen}},\ and\ \citenamefont
  {Sachdev}}]{witczak-krempa2014}%
  \BibitemOpen
  \bibfield  {author} {\bibinfo {author} {\bibfnamefont {W.}~\bibnamefont
  {Witczak-Krempa}}, \bibinfo {author} {\bibfnamefont {E.~S.}\ \bibnamefont
  {{S{\o}rensen}}}, \ and\ \bibinfo {author} {\bibfnamefont {S.}~\bibnamefont
  {Sachdev}},\ }\href {\doibase 10.1038/nphys2913} {\bibfield  {journal}
  {\bibinfo  {journal} {Nature Phys.}\ }\textbf {\bibinfo {volume} {10}},\
  \bibinfo {pages} {361} (\bibinfo {year} {2014})}\BibitemShut {NoStop}%
\bibitem [{\citenamefont {Katz}\ \emph {et~al.}(2014)\citenamefont {Katz},
  \citenamefont {Sachdev}, \citenamefont {{S{\o}rensen}},\ and\ \citenamefont
  {Witczak-Krempa}}]{katz2014}%
  \BibitemOpen
  \bibfield  {author} {\bibinfo {author} {\bibfnamefont {E.}~\bibnamefont
  {Katz}}, \bibinfo {author} {\bibfnamefont {S.}~\bibnamefont {Sachdev}},
  \bibinfo {author} {\bibfnamefont {E.~S.}\ \bibnamefont {{S{\o}rensen}}}, \
  and\ \bibinfo {author} {\bibfnamefont {W.}~\bibnamefont {Witczak-Krempa}},\
  }\href {\doibase 10.1103/PhysRevB.90.245109} {\bibfield  {journal} {\bibinfo
  {journal} {Phys. Rev. B}\ }\textbf {\bibinfo {volume} {90}},\ \bibinfo
  {pages} {245109} (\bibinfo {year} {2014})}\BibitemShut {NoStop}%
\bibitem [{\citenamefont {Gazit}\ \emph {et~al.}(2014)\citenamefont {Gazit},
  \citenamefont {Podolsky},\ and\ \citenamefont {Auerbach}}]{gazit2014}%
  \BibitemOpen
  \bibfield  {author} {\bibinfo {author} {\bibfnamefont {S.}~\bibnamefont
  {Gazit}}, \bibinfo {author} {\bibfnamefont {D.}~\bibnamefont {Podolsky}}, \
  and\ \bibinfo {author} {\bibfnamefont {A.}~\bibnamefont {Auerbach}},\ }\href
  {\doibase 10.1103/PhysRevLett.113.240601} {\bibfield  {journal} {\bibinfo
  {journal} {Phys. Rev. Lett.}\ }\textbf {\bibinfo {volume} {113}},\ \bibinfo
  {pages} {240601} (\bibinfo {year} {2014})}\BibitemShut {NoStop}%
\bibitem [{\citenamefont {Kos}\ \emph {et~al.}(2015)\citenamefont {Kos},
  \citenamefont {Poland}, \citenamefont {Simmons-Duffin},\ and\ \citenamefont
  {Vichi}}]{kos2015}%
  \BibitemOpen
  \bibfield  {author} {\bibinfo {author} {\bibfnamefont {F.}~\bibnamefont
  {Kos}}, \bibinfo {author} {\bibfnamefont {D.}~\bibnamefont {Poland}},
  \bibinfo {author} {\bibfnamefont {D.}~\bibnamefont {Simmons-Duffin}}, \ and\
  \bibinfo {author} {\bibfnamefont {A.}~\bibnamefont {Vichi}},\ }\href
  {\doibase 10.1007/JHEP11(2015)106} {\bibfield  {journal} {\bibinfo  {journal}
  {JHEP}\ }\textbf {\bibinfo {volume} {11}},\ \bibinfo {pages} {106} (\bibinfo
  {year} {2015})}\BibitemShut {NoStop}%
\bibitem [{\citenamefont {Cha}\ \emph {et~al.}(1991)\citenamefont {Cha},
  \citenamefont {Fisher}, \citenamefont {Girvin}, \citenamefont {Wallin},\ and\
  \citenamefont {Young}}]{cha1991}%
  \BibitemOpen
  \bibfield  {author} {\bibinfo {author} {\bibfnamefont {M.-C.}\ \bibnamefont
  {Cha}}, \bibinfo {author} {\bibfnamefont {M.~P.~A.}\ \bibnamefont {Fisher}},
  \bibinfo {author} {\bibfnamefont {S.~M.}\ \bibnamefont {Girvin}}, \bibinfo
  {author} {\bibfnamefont {M.}~\bibnamefont {Wallin}}, \ and\ \bibinfo {author}
  {\bibfnamefont {A.~P.}\ \bibnamefont {Young}},\ }\href {\doibase
  10.1103/PhysRevB.44.6883} {\bibfield  {journal} {\bibinfo  {journal} {Phys.
  Rev. B}\ }\textbf {\bibinfo {volume} {44}},\ \bibinfo {pages} {6883}
  (\bibinfo {year} {1991})}\BibitemShut {NoStop}%
\bibitem [{\citenamefont {Fazio}\ and\ \citenamefont
  {Zappal\`{a}}(1996)}]{fazio1996}%
  \BibitemOpen
  \bibfield  {author} {\bibinfo {author} {\bibfnamefont {R.}~\bibnamefont
  {Fazio}}\ and\ \bibinfo {author} {\bibfnamefont {D.}~\bibnamefont
  {Zappal\`{a}}},\ }\href {\doibase 10.1103/PhysRevB.53.R8883} {\bibfield
  {journal} {\bibinfo  {journal} {Phys. Rev. B}\ }\textbf {\bibinfo {volume}
  {53}},\ \bibinfo {pages} {R8883} (\bibinfo {year} {1996})}\BibitemShut
  {NoStop}%
\bibitem [{\citenamefont {{\v{S}makov}}\ and\ \citenamefont
  {{S{\o}rensen}}(2005)}]{smakov2005}%
  \BibitemOpen
  \bibfield  {author} {\bibinfo {author} {\bibfnamefont {J.}~\bibnamefont
  {{\v{S}makov}}}\ and\ \bibinfo {author} {\bibfnamefont {E.}~\bibnamefont
  {{S{\o}rensen}}},\ }\href {\doibase 10.1103/PhysRevLett.95.180603} {\bibfield
   {journal} {\bibinfo  {journal} {Phys. Rev. Lett.}\ }\textbf {\bibinfo
  {volume} {95}},\ \bibinfo {pages} {180603} (\bibinfo {year}
  {2005})}\BibitemShut {NoStop}%
\bibitem [{\citenamefont {Vidal}\ \emph {et~al.}(2003)\citenamefont {Vidal},
  \citenamefont {Latorre}, \citenamefont {Rico},\ and\ \citenamefont
  {Kitaev}}]{vidal2003}%
  \BibitemOpen
  \bibfield  {author} {\bibinfo {author} {\bibfnamefont {G.}~\bibnamefont
  {Vidal}}, \bibinfo {author} {\bibfnamefont {J.~I.}\ \bibnamefont {Latorre}},
  \bibinfo {author} {\bibfnamefont {E.}~\bibnamefont {Rico}}, \ and\ \bibinfo
  {author} {\bibfnamefont {A.}~\bibnamefont {Kitaev}},\ }\href {\doibase
  10.1103/PhysRevLett.90.227902} {\bibfield  {journal} {\bibinfo  {journal}
  {Phys. Rev. Lett.}\ }\textbf {\bibinfo {volume} {90}},\ \bibinfo {pages}
  {227902} (\bibinfo {year} {2003})}\BibitemShut {NoStop}%
\bibitem [{\citenamefont {Calabrese}\ and\ \citenamefont
  {Cardy}(2004)}]{calabrese2004}%
  \BibitemOpen
  \bibfield  {author} {\bibinfo {author} {\bibfnamefont {P.}~\bibnamefont
  {Calabrese}}\ and\ \bibinfo {author} {\bibfnamefont {J.}~\bibnamefont
  {Cardy}},\ }\href {\doibase 10.1088/1742-5468/2004/06/P06002} {\bibfield
  {journal} {\bibinfo  {journal} {J. Stat. Mech.}\ }\textbf {\bibinfo {volume}
  {2004}},\ \bibinfo {pages} {P06002} (\bibinfo {year} {2004})}\BibitemShut
  {NoStop}%
\bibitem [{\citenamefont {Stoudenmire}\ \emph {et~al.}(2014)\citenamefont
  {Stoudenmire}, \citenamefont {Gustainis}, \citenamefont {Johal},
  \citenamefont {Wessel},\ and\ \citenamefont {Melko}}]{stoudenmire2014}%
  \BibitemOpen
  \bibfield  {author} {\bibinfo {author} {\bibfnamefont {E.~M.}\ \bibnamefont
  {Stoudenmire}}, \bibinfo {author} {\bibfnamefont {P.}~\bibnamefont
  {Gustainis}}, \bibinfo {author} {\bibfnamefont {R.}~\bibnamefont {Johal}},
  \bibinfo {author} {\bibfnamefont {S.}~\bibnamefont {Wessel}}, \ and\ \bibinfo
  {author} {\bibfnamefont {R.~G.}\ \bibnamefont {Melko}},\ }\href {\doibase
  10.1103/PhysRevB.90.235106} {\bibfield  {journal} {\bibinfo  {journal} {Phys.
  Rev. B}\ }\textbf {\bibinfo {volume} {90}},\ \bibinfo {pages} {235106}
  (\bibinfo {year} {2014})}\BibitemShut {NoStop}%
\bibitem [{\citenamefont {Kallin}\ \emph {et~al.}(2014)\citenamefont {Kallin},
  \citenamefont {Stoudenmire}, \citenamefont {Fendley}, \citenamefont {Singh},\
  and\ \citenamefont {Melko}}]{kallin2014}%
  \BibitemOpen
  \bibfield  {author} {\bibinfo {author} {\bibfnamefont {A.~B.}\ \bibnamefont
  {Kallin}}, \bibinfo {author} {\bibfnamefont {E.~M.}\ \bibnamefont
  {Stoudenmire}}, \bibinfo {author} {\bibfnamefont {P.}~\bibnamefont
  {Fendley}}, \bibinfo {author} {\bibfnamefont {R.~R.~P.}\ \bibnamefont
  {Singh}}, \ and\ \bibinfo {author} {\bibfnamefont {R.~G.}\ \bibnamefont
  {Melko}},\ }\href {\doibase 10.1088/1742-5468/2014/06/P06009} {\bibfield
  {journal} {\bibinfo  {journal} {J. Stat. Mech.}\ }\textbf {\bibinfo {volume}
  {2014}},\ \bibinfo {pages} {P06009} (\bibinfo {year} {2014})}\BibitemShut
  {NoStop}%
\bibitem [{\citenamefont {Helmes}\ and\ \citenamefont
  {Wessel}(2015)}]{wessel15}%
  \BibitemOpen
  \bibfield  {author} {\bibinfo {author} {\bibfnamefont {J.}~\bibnamefont
  {Helmes}}\ and\ \bibinfo {author} {\bibfnamefont {S.}~\bibnamefont
  {Wessel}},\ }\href {\doibase 10.1103/PhysRevB.92.125120} {\bibfield
  {journal} {\bibinfo  {journal} {Phys. Rev. B}\ }\textbf {\bibinfo {volume}
  {92}},\ \bibinfo {pages} {125120} (\bibinfo {year} {2015})}\BibitemShut
  {NoStop}%
\bibitem [{\citenamefont {Bueno}\ \emph
  {et~al.}(2015{\natexlab{a}})\citenamefont {Bueno}, \citenamefont {Myers},\
  and\ \citenamefont {Witczak-Krempa}}]{bueno2015}%
  \BibitemOpen
  \bibfield  {author} {\bibinfo {author} {\bibfnamefont {P.}~\bibnamefont
  {Bueno}}, \bibinfo {author} {\bibfnamefont {R.~C.}\ \bibnamefont {Myers}}, \
  and\ \bibinfo {author} {\bibfnamefont {W.}~\bibnamefont {Witczak-Krempa}},\
  }\href {\doibase 10.1103/PhysRevLett.115.021602} {\bibfield  {journal}
  {\bibinfo  {journal} {Phys. Rev. Lett.}\ }\textbf {\bibinfo {volume} {115}},\
  \bibinfo {pages} {021602} (\bibinfo {year} {2015}{\natexlab{a}})}\BibitemShut
  {NoStop}%
\bibitem [{\citenamefont {Faulkner}\ \emph {et~al.}(2015)\citenamefont
  {Faulkner}, \citenamefont {Leigh},\ and\ \citenamefont
  {Parrikar}}]{faulkner15}%
  \BibitemOpen
  \bibfield  {author} {\bibinfo {author} {\bibfnamefont {T.}~\bibnamefont
  {Faulkner}}, \bibinfo {author} {\bibfnamefont {R.~G.}\ \bibnamefont {Leigh}},
  \ and\ \bibinfo {author} {\bibfnamefont {O.}~\bibnamefont {Parrikar}},\
  }\href {http://arxiv.org/abs/1511.05179} {\bibfield  {journal} {\bibinfo
  {journal} {arXiv:1511.05179}\ } (\bibinfo {year} {2015})}\BibitemShut
  {NoStop}%
\bibitem [{\citenamefont {{Bueno}}\ and\ \citenamefont
  {{Myers}}(2015)}]{bueno1505}%
  \BibitemOpen
  \bibfield  {author} {\bibinfo {author} {\bibfnamefont {P.}~\bibnamefont
  {{Bueno}}}\ and\ \bibinfo {author} {\bibfnamefont {R.~C.}\ \bibnamefont
  {{Myers}}},\ }\href {\doibase 10.1007/JHEP08(2015)068} {\bibfield  {journal}
  {\bibinfo  {journal} {JHEP}\ }\textbf {\bibinfo {volume} {1508}},\ \bibinfo
  {pages} {068} (\bibinfo {year} {2015})}\BibitemShut {NoStop}%
\bibitem [{\citenamefont {{Bueno}}\ and\ \citenamefont
  {{Witczak-Krempa}}(2016)}]{bound}%
  \BibitemOpen
  \bibfield  {author} {\bibinfo {author} {\bibfnamefont {P.}~\bibnamefont
  {{Bueno}}}\ and\ \bibinfo {author} {\bibfnamefont {W.}~\bibnamefont
  {{Witczak-Krempa}}},\ }\href {\doibase 10.1103/PhysRevB.93.045131} {\bibfield
   {journal} {\bibinfo  {journal} {\prb}\ }\textbf {\bibinfo {volume} {93}},\
  \bibinfo {eid} {045131} (\bibinfo {year} {2016})}\BibitemShut {NoStop}%
\bibitem [{\citenamefont {Bobev}\ \emph {et~al.}(2015)\citenamefont {Bobev},
  \citenamefont {El-Showk}, \citenamefont {{Maz\'{a}\v{c}}},\ and\
  \citenamefont {Paulos}}]{bobev2015}%
  \BibitemOpen
  \bibfield  {author} {\bibinfo {author} {\bibfnamefont {N.}~\bibnamefont
  {Bobev}}, \bibinfo {author} {\bibfnamefont {S.}~\bibnamefont {El-Showk}},
  \bibinfo {author} {\bibfnamefont {D.}~\bibnamefont {{Maz\'{a}\v{c}}}}, \ and\
  \bibinfo {author} {\bibfnamefont {M.~F.}\ \bibnamefont {Paulos}},\ }\href
  {\doibase 10.1103/PhysRevLett.115.051601} {\bibfield  {journal} {\bibinfo
  {journal} {Phys. Rev. Lett.}\ }\textbf {\bibinfo {volume} {115}},\ \bibinfo
  {pages} {051601} (\bibinfo {year} {2015})}\BibitemShut {NoStop}%
\bibitem [{\citenamefont {{Gulotta}}\ \emph {et~al.}(2011)\citenamefont
  {{Gulotta}}, \citenamefont {{Herzog}},\ and\ \citenamefont
  {{Kaminski}}}]{herzog2011}%
  \BibitemOpen
  \bibfield  {author} {\bibinfo {author} {\bibfnamefont {D.~R.}\ \bibnamefont
  {{Gulotta}}}, \bibinfo {author} {\bibfnamefont {C.~P.}\ \bibnamefont
  {{Herzog}}}, \ and\ \bibinfo {author} {\bibfnamefont {M.}~\bibnamefont
  {{Kaminski}}},\ }\href {\doibase 10.1007/JHEP01(2011)148} {\bibfield
  {journal} {\bibinfo  {journal} {JHEP}\ }\textbf {\bibinfo {volume} {1101}},\
  \bibinfo {eid} {148} (\bibinfo {year} {2011})}\BibitemShut {NoStop}%
\bibitem [{\citenamefont {{Witczak-Krempa}}\ and\ \citenamefont
  {{Sachdev}}(2012)}]{will_qnm}%
  \BibitemOpen
  \bibfield  {author} {\bibinfo {author} {\bibfnamefont {W.}~\bibnamefont
  {{Witczak-Krempa}}}\ and\ \bibinfo {author} {\bibfnamefont {S.}~\bibnamefont
  {{Sachdev}}},\ }\href {\doibase 10.1103/PhysRevB.86.235115} {\bibfield
  {journal} {\bibinfo  {journal} {\prb}\ }\textbf {\bibinfo {volume} {86}},\
  \bibinfo {eid} {235115} (\bibinfo {year} {2012})}\BibitemShut {NoStop}%
\bibitem [{\citenamefont {Harris}(1974)}]{harris1974}%
  \BibitemOpen
  \bibfield  {author} {\bibinfo {author} {\bibfnamefont {A.~B.}\ \bibnamefont
  {Harris}},\ }\href {\doibase 10.1088/0022-3719/7/9/009} {\bibfield  {journal}
  {\bibinfo  {journal} {J. Phys. C}\ }\textbf {\bibinfo {volume} {7}},\
  \bibinfo {pages} {1671} (\bibinfo {year} {1974})}\BibitemShut {NoStop}%
\bibitem [{\citenamefont {Nandkishore}\ \emph {et~al.}(2013)\citenamefont
  {Nandkishore}, \citenamefont {Maciejko}, \citenamefont {Huse},\ and\
  \citenamefont {Sondhi}}]{nandkishore2013}%
  \BibitemOpen
  \bibfield  {author} {\bibinfo {author} {\bibfnamefont {R.}~\bibnamefont
  {Nandkishore}}, \bibinfo {author} {\bibfnamefont {J.}~\bibnamefont
  {Maciejko}}, \bibinfo {author} {\bibfnamefont {D.~A.}\ \bibnamefont {Huse}},
  \ and\ \bibinfo {author} {\bibfnamefont {S.~L.}\ \bibnamefont {Sondhi}},\
  }\href {\doibase 10.1103/PhysRevB.87.174511} {\bibfield  {journal} {\bibinfo
  {journal} {Phys. Rev. B}\ }\textbf {\bibinfo {volume} {87}},\ \bibinfo
  {pages} {174511} (\bibinfo {year} {2013})}\BibitemShut {NoStop}%
\bibitem [{\citenamefont {Bueno}\ \emph
  {et~al.}(2015{\natexlab{b}})\citenamefont {Bueno}, \citenamefont {Myers},\
  and\ \citenamefont {Witczak-Krempa}}]{twist}%
  \BibitemOpen
  \bibfield  {author} {\bibinfo {author} {\bibfnamefont {P.}~\bibnamefont
  {Bueno}}, \bibinfo {author} {\bibfnamefont {R.~C.}\ \bibnamefont {Myers}}, \
  and\ \bibinfo {author} {\bibfnamefont {W.}~\bibnamefont {Witczak-Krempa}},\
  }\href {\doibase 10.1007/JHEP09(2015)091} {\bibfield  {journal} {\bibinfo
  {journal} {JHEP}\ }\textbf {\bibinfo {volume} {1509}},\ \bibinfo {pages}
  {091} (\bibinfo {year} {2015}{\natexlab{b}})}\BibitemShut {NoStop}%
\end{thebibliography}%


\begin{thebibliography}{12}%
\makeatletter
\providecommand \@ifxundefined [1]{%
 \@ifx{#1\undefined}
}%
\providecommand \@ifnum [1]{%
 \ifnum #1\expandafter \@firstoftwo
 \else \expandafter \@secondoftwo
 \fi
}%
\providecommand \@ifx [1]{%
 \ifx #1\expandafter \@firstoftwo
 \else \expandafter \@secondoftwo
 \fi
}%
\providecommand \natexlab [1]{#1}%
\providecommand \enquote  [1]{``#1''}%
\providecommand \bibnamefont  [1]{#1}%
\providecommand \bibfnamefont [1]{#1}%
\providecommand \citenamefont [1]{#1}%
\providecommand \href@noop [0]{\@secondoftwo}%
\providecommand \href [0]{\begingroup \@sanitize@url \@href}%
\providecommand \@href[1]{\@@startlink{#1}\@@href}%
\providecommand \@@href[1]{\endgroup#1\@@endlink}%
\providecommand \@sanitize@url [0]{\catcode `\\12\catcode `\$12\catcode
  `\&12\catcode `\#12\catcode `\^12\catcode `\_12\catcode `\%12\relax}%
\providecommand \@@startlink[1]{}%
\providecommand \@@endlink[0]{}%
\providecommand \url  [0]{\begingroup\@sanitize@url \@url }%
\providecommand \@url [1]{\endgroup\@href {#1}{\urlprefix }}%
\providecommand \urlprefix  [0]{URL }%
\providecommand \Eprint [0]{\href }%
\providecommand \doibase [0]{http://dx.doi.org/}%
\providecommand \selectlanguage [0]{\@gobble}%
\providecommand \bibinfo  [0]{\@secondoftwo}%
\providecommand \bibfield  [0]{\@secondoftwo}%
\providecommand \translation [1]{[#1]}%
\providecommand \BibitemOpen [0]{}%
\providecommand \bibitemStop [0]{}%
\providecommand \bibitemNoStop [0]{.\EOS\space}%
\providecommand \EOS [0]{\spacefactor3000\relax}%
\providecommand \BibitemShut  [1]{\csname bibitem#1\endcsname}%
\let\auto@bib@innerbib\@empty
\bibitem [{\citenamefont {Dumitrescu}\ and\ \citenamefont
  {Seiberg}(2011)}]{dumitrescu2011}%
  \BibitemOpen
  \bibfield  {author} {\bibinfo {author} {\bibfnamefont {T.~T.}\ \bibnamefont
  {Dumitrescu}}\ and\ \bibinfo {author} {\bibfnamefont {N.}~\bibnamefont
  {Seiberg}},\ }\bibfield  {title} {\enquote {\bibinfo {title} {Supercurrents
  and brane currents in diverse dimensions},}\ }\href {\doibase
  10.1007/JHEP07(2011)095} {\bibfield  {journal} {\bibinfo  {journal} {JHEP}\
  }\textbf {\bibinfo {volume} {1107}},\ \bibinfo {pages} {095} (\bibinfo {year}
  {2011})}\BibitemShut {NoStop}%
\bibitem [{\citenamefont {Buchbinder}\ \emph {et~al.}(2015)\citenamefont
  {Buchbinder}, \citenamefont {Kuzenko},\ and\ \citenamefont
  {Samsonov}}]{buchbinder2015}%
  \BibitemOpen
  \bibfield  {author} {\bibinfo {author} {\bibfnamefont {Evgeny~I.}\
  \bibnamefont {Buchbinder}}, \bibinfo {author} {\bibfnamefont {Sergei~M.}\
  \bibnamefont {Kuzenko}}, \ and\ \bibinfo {author} {\bibfnamefont {Igor~B.}\
  \bibnamefont {Samsonov}},\ }\bibfield  {title} {\enquote {\bibinfo {title}
  {Superconformal field theory in three dimensions: correlation functions of
  conserved currents},}\ }\href {\doibase 10.1007/JHEP06(2015)138} {\bibfield
  {journal} {\bibinfo  {journal} {JHEP}\ }\textbf {\bibinfo {volume} {1506}},\
  \bibinfo {pages} {138} (\bibinfo {year} {2015})}\BibitemShut {NoStop}%
\bibitem [{\citenamefont {Osborn}\ and\ \citenamefont
  {Petkou}(1994)}]{osborn1994}%
  \BibitemOpen
  \bibfield  {author} {\bibinfo {author} {\bibfnamefont {H.}~\bibnamefont
  {Osborn}}\ and\ \bibinfo {author} {\bibfnamefont {A.}~\bibnamefont
  {Petkou}},\ }\bibfield  {title} {\enquote {\bibinfo {title} {Implications of
  conformal invariance in field theories for general dimensions},}\ }\href
  {\doibase 10.1006/aphy.1994.1045} {\bibfield  {journal} {\bibinfo  {journal}
  {Ann. Phys.}\ }\textbf {\bibinfo {volume} {231}},\ \bibinfo {pages} {311}
  (\bibinfo {year} {1994})}\BibitemShut {NoStop}%
\bibitem [{\citenamefont {Barnes}\ \emph {et~al.}(2005)\citenamefont {Barnes},
  \citenamefont {Gorbatov}, \citenamefont {Intriligator}, \citenamefont
  {Sudano},\ and\ \citenamefont {Wright}}]{barnes2005}%
  \BibitemOpen
  \bibfield  {author} {\bibinfo {author} {\bibfnamefont {E.}~\bibnamefont
  {Barnes}}, \bibinfo {author} {\bibfnamefont {E.}~\bibnamefont {Gorbatov}},
  \bibinfo {author} {\bibfnamefont {K.}~\bibnamefont {Intriligator}}, \bibinfo
  {author} {\bibfnamefont {M.}~\bibnamefont {Sudano}}, \ and\ \bibinfo {author}
  {\bibfnamefont {J.}~\bibnamefont {Wright}},\ }\bibfield  {title} {\enquote
  {\bibinfo {title} {The exact superconformal {$R$}-symmetry minimizes
  $\tau_{RR}$},}\ }\href {\doibase 10.1016/j.nuclphysb.2005.10.003} {\bibfield
  {journal} {\bibinfo  {journal} {Nucl. Phys. B}\ }\textbf {\bibinfo {volume}
  {730}},\ \bibinfo {pages} {210} (\bibinfo {year} {2005})}\BibitemShut
  {NoStop}%
\bibitem [{\citenamefont {Nishioka}\ and\ \citenamefont
  {Yonekura}(2013)}]{nishioka2013}%
  \BibitemOpen
  \bibfield  {author} {\bibinfo {author} {\bibfnamefont {T.}~\bibnamefont
  {Nishioka}}\ and\ \bibinfo {author} {\bibfnamefont {K.}~\bibnamefont
  {Yonekura}},\ }\bibfield  {title} {\enquote {\bibinfo {title} {On {RG} flow
  of {$\tau_{RR}$} for supersymmetric field theories in three-dimensions},}\
  }\href {\doibase 10.1007/JHEP05(2013)165} {\bibfield  {journal} {\bibinfo
  {journal} {JHEP}\ }\textbf {\bibinfo {volume} {1305}},\ \bibinfo {pages}
  {165} (\bibinfo {year} {2013})}\BibitemShut {NoStop}%
\bibitem [{\citenamefont {Imamura}(2011)}]{imamura2011}%
  \BibitemOpen
  \bibfield  {author} {\bibinfo {author} {\bibfnamefont {Y.}~\bibnamefont
  {Imamura}},\ }\bibfield  {title} {\enquote {\bibinfo {title} {Relation
  between the 4d superconformal index and the {$S^3$} partition function},}\
  }\href {\doibase 10.1007/JHEP09(2011)133} {\bibfield  {journal} {\bibinfo
  {journal} {JHEP}\ }\textbf {\bibinfo {volume} {1109}},\ \bibinfo {pages}
  {133} (\bibinfo {year} {2011})}\BibitemShut {NoStop}%
\bibitem [{\citenamefont {Imamura}\ and\ \citenamefont
  {Yokoyama}(2012)}]{imamura2012}%
  \BibitemOpen
  \bibfield  {author} {\bibinfo {author} {\bibfnamefont {Y.}~\bibnamefont
  {Imamura}}\ and\ \bibinfo {author} {\bibfnamefont {D.}~\bibnamefont
  {Yokoyama}},\ }\bibfield  {title} {\enquote {\bibinfo {title}
  {$\mathcal{N}=2$ supersymmetric theories on squashed three-sphere},}\ }\href
  {\doibase 10.1103/PhysRevD.85.025015} {\bibfield  {journal} {\bibinfo
  {journal} {Phys. Rev. D}\ }\textbf {\bibinfo {volume} {85}},\ \bibinfo
  {pages} {025015} (\bibinfo {year} {2012})}\BibitemShut {NoStop}%
\bibitem [{\citenamefont {Aharony}\ \emph {et~al.}(1997)\citenamefont
  {Aharony}, \citenamefont {Hanany}, \citenamefont {Intriligator},
  \citenamefont {Seiberg},\ and\ \citenamefont {Strassler}}]{aharony1997}%
  \BibitemOpen
  \bibfield  {author} {\bibinfo {author} {\bibfnamefont {O.}~\bibnamefont
  {Aharony}}, \bibinfo {author} {\bibfnamefont {A.}~\bibnamefont {Hanany}},
  \bibinfo {author} {\bibfnamefont {K.}~\bibnamefont {Intriligator}}, \bibinfo
  {author} {\bibfnamefont {N.}~\bibnamefont {Seiberg}}, \ and\ \bibinfo
  {author} {\bibfnamefont {M.~J.}\ \bibnamefont {Strassler}},\ }\bibfield
  {title} {\enquote {\bibinfo {title} {Aspects of {$\mathcal{N}=2$}
  supersymmetric gauge theories in three dimensions},}\ }\href {\doibase
  10.1016/S0550-3213(97)00323-4} {\bibfield  {journal} {\bibinfo  {journal}
  {Nucl. Phys. B}\ }\textbf {\bibinfo {volume} {499}},\ \bibinfo {pages} {67}
  (\bibinfo {year} {1997})}\BibitemShut {NoStop}%
\bibitem [{\citenamefont {Bobev}\ \emph {et~al.}(2015)\citenamefont {Bobev},
  \citenamefont {El-Showk}, \citenamefont {{Maz\'{a}\v{c}}},\ and\
  \citenamefont {Paulos}}]{bobev2015}%
  \BibitemOpen
  \bibfield  {author} {\bibinfo {author} {\bibfnamefont {Nikolay}\ \bibnamefont
  {Bobev}}, \bibinfo {author} {\bibfnamefont {Sheer}\ \bibnamefont {El-Showk}},
  \bibinfo {author} {\bibfnamefont {Dalimil}\ \bibnamefont {{Maz\'{a}\v{c}}}},
  \ and\ \bibinfo {author} {\bibfnamefont {Miguel~F.}\ \bibnamefont {Paulos}},\
  }\bibfield  {title} {\enquote {\bibinfo {title} {Bootstrapping the {Three}
  {Dimensional} {Supersymmetric} {Ising} {Model}},}\ }\href {\doibase
  10.1103/PhysRevLett.115.051601} {\bibfield  {journal} {\bibinfo  {journal}
  {Phys. Rev. Lett.}\ }\textbf {\bibinfo {volume} {115}},\ \bibinfo {pages}
  {051601} (\bibinfo {year} {2015})}\BibitemShut {NoStop}%
\bibitem [{\citenamefont {{Hofman}}\ and\ \citenamefont
  {{Maldacena}}(2008)}]{hofman08}%
  \BibitemOpen
  \bibfield  {author} {\bibinfo {author} {\bibfnamefont {D.~M.}\ \bibnamefont
  {{Hofman}}}\ and\ \bibinfo {author} {\bibfnamefont {J.}~\bibnamefont
  {{Maldacena}}},\ }\bibfield  {title} {\enquote {\bibinfo {title} {{Conformal
  collider physics: energy and charge correlations}},}\ }\href {\doibase
  10.1088/1126-6708/2008/05/012} {\bibfield  {journal} {\bibinfo  {journal}
  {JHEP}\ }\textbf {\bibinfo {volume} {0805}},\ \bibinfo {eid} {012} (\bibinfo
  {year} {2008})}\BibitemShut {NoStop}%
\bibitem [{\citenamefont {{Myers}}\ \emph {et~al.}(2011)\citenamefont
  {{Myers}}, \citenamefont {{Sachdev}},\ and\ \citenamefont {{Singh}}}]{ajay}%
  \BibitemOpen
  \bibfield  {author} {\bibinfo {author} {\bibfnamefont {R.~C.}\ \bibnamefont
  {{Myers}}}, \bibinfo {author} {\bibfnamefont {S.}~\bibnamefont {{Sachdev}}},
  \ and\ \bibinfo {author} {\bibfnamefont {A.}~\bibnamefont {{Singh}}},\
  }\bibfield  {title} {\enquote {\bibinfo {title} {{Holographic quantum
  critical transport without self-duality}},}\ }\href {\doibase
  10.1103/PhysRevD.83.066017} {\bibfield  {journal} {\bibinfo  {journal}
  {\prd}\ }\textbf {\bibinfo {volume} {83}},\ \bibinfo {eid} {066017} (\bibinfo
  {year} {2011})}\BibitemShut {NoStop}%
\bibitem [{\citenamefont {Chowdhury}\ \emph {et~al.}(2013)\citenamefont
  {Chowdhury}, \citenamefont {Raju}, \citenamefont {Sachdev}, \citenamefont
  {Singh},\ and\ \citenamefont {Strack}}]{chowdhury2013}%
  \BibitemOpen
  \bibfield  {author} {\bibinfo {author} {\bibfnamefont {D.}~\bibnamefont
  {Chowdhury}}, \bibinfo {author} {\bibfnamefont {S.}~\bibnamefont {Raju}},
  \bibinfo {author} {\bibfnamefont {S.}~\bibnamefont {Sachdev}}, \bibinfo
  {author} {\bibfnamefont {A.}~\bibnamefont {Singh}}, \ and\ \bibinfo {author}
  {\bibfnamefont {P.}~\bibnamefont {Strack}},\ }\bibfield  {title} {\enquote
  {\bibinfo {title} {Multipoint correlators of conformal field theories:
  {Implications} for quantum critical transport},}\ }\href {\doibase
  10.1103/PhysRevB.87.085138} {\bibfield  {journal} {\bibinfo  {journal} {Phy.
  Rev. B}\ }\textbf {\bibinfo {volume} {87}},\ \bibinfo {pages} {085138}
  (\bibinfo {year} {2013})}\BibitemShut {NoStop}%
\end{thebibliography}%

\end{document}


\newcommand{\tr}{\mathop{\mathrm{tr}}}
\newcommand{\bsigma}{\boldsymbol{\sigma}}
\newcommand{\re}{\mathop{\mathrm{Re}}}
\newcommand{\im}{\mathop{\mathrm{Im}}}
\renewcommand{\b}[1]{{\boldsymbol{#1}}}
\newcommand{\diag}{\mathrm{diag}}
\newcommand{\sign}{\mathrm{sign}}
\newcommand{\sgn}{\mathop{\mathrm{sgn}}}
\renewcommand{\c}[1]{\mathcal{#1}}

\newcommand{\mb}{\bm}
\newcommand{\ua}{\uparrow}
\newcommand{\da}{\downarrow}
\newcommand{\ra}{\rightarrow}
\newcommand{\la}{\leftarrow}
\newcommand{\mc}{\mathcal}
\newcommand{\bs}{\boldsymbol}
\newcommand{\lra}{\leftrightarrow}
\newcommand{\nn}{\nonumber}
\newcommand{\half}{{\textstyle{\frac{1}{2}}}}
\newcommand{\mf}{\mathfrak}
\newcommand{\MF}{\text{MF}}
\newcommand{\IR}{\text{IR}}
\newcommand{\UV}{\text{UV}}

\DeclareGraphicsExtensions{.png}

\renewcommand{\thetable}{S\Roman{table}}
\renewcommand{\thefigure}{S\arabic{figure}}
\renewcommand{\theequation}{S\arabic{equation}}

\title{Supplemental material for\\ ``Optical conductivity of topological surface states with emergent supersymmetry''}
 
\author{William Witczak-Krempa}
\affiliation{Department of Physics, Harvard University, Cambridge, Massachusetts 02138, USA}

\author{Joseph Maciejko}
\affiliation{Department of Physics, University of Alberta, Edmonton, Alberta T6G 2E1, Canada}
\affiliation{Theoretical Physics Institute, University of Alberta, Edmonton, Alberta T6G 2E1, Canada}
\affiliation{Canadian Institute for Advanced Research, Toronto, Ontario M5G 1Z8, Canada}

\date{\today\\
\vspace{0.6in}}


\maketitle 
\tableofcontents
 
\section{Two-point functions of $\mathcal{N}=2$ SCFTs in 2+1 dimensions}

In supersymmetric theories, fields are grouped into different supermultiplets according to how they transform under the supersymmetry algebra. For (2+1)-dimensional theories with $\mathcal{N}=2$ supersymmetry, the $R$-current $R_\mu$ and the stress tensor $T_{\mu\nu}$ are in the same supermultiplet, the supercurrent supermultiplet. For the (2+1)D Wess-Zumino model we are considering, the $R$-current is simply proportional to the physical $U(1)$ current $J_\mu=\bar{\psi}\gamma_\mu\psi+i(\phi^*\partial_\mu\phi-\mathrm{c.c.})$. In the superspace formalism, one associates to each supermultiplet a superfield which contains the various components of the supermultiplet. For the supercurrent supermultiplet, the superfield $\mathcal{J}_\mu$ is~\cite{dumitrescu2011}
\begin{align}\label{superJ}
\mathcal{J}_\mu=sJ_\mu-(\theta\gamma^\nu\bar{\theta})2T_{\nu\mu}+\dotsb,
\end{align}
where $s$ is the proportionality constant between $R_\mu$ and $J_\mu$, $(\dotsb)$ denotes all the components other than the $U(1)$ current and the stress tensor, the $\gamma^\nu$ ($\nu=0,1,2$) 
are $2\times 2$ gamma matrices, and $\theta,\bar{\theta}$ are Grassmann-valued two-component spinors. 
In this Supplemental Material we consider Minkowski spacetime described by the metric tensor $\eta_{\mu\nu}=\diag(-1,1,1)$, i.e., a Lorentzian metric with signature $(-++)$, but results in imaginary time (Euclidean spacetime) can be obtained simply by replacing $\eta_{\mu\nu}$ with the Kronecker delta $\delta_{\mu\nu}$. Lorentz indices $\mu,\nu,\ldots$ are lowered (raised) with the metric tensor $\eta_{\mu\nu}$ ($\eta^{\mu\nu})$, while spinorial indices $\alpha,\beta,\ldots$ are lowered (raised) with the antisymmetric tensor $\varepsilon_{\alpha\beta}$ ($\varepsilon^{\alpha\beta}$), where we define
\begin{align}
\varepsilon_{\alpha\beta}\equiv\left(\begin{array}{cc}
0 & -1 \\
1 & 0
\end{array}\right)=-i\sigma_2,\hspace{5mm}
\varepsilon^{\alpha\beta}\equiv\left(\begin{array}{cc}
0 & 1 \\
-1 & 0
\end{array}\right)=i\sigma_2\,,
\end{align}
where $\sigma_1,\sigma_2,\sigma_3$ are the standard Pauli matrices. Because $\varepsilon_{\alpha\beta}$ and $\varepsilon^{\alpha\beta}$ are antisymmetric tensors, one must be careful to use the second index of the pair when lowering and raising spinorial indices,
\begin{align}
\theta_\alpha=\varepsilon_{\alpha\beta}\theta^\beta,\hspace{5mm}
\theta^\alpha=\varepsilon^{\alpha\beta}\theta_\beta\,.
\end{align}
The gamma matrices are defined as
\begin{align}\label{DefGammaMu}
(\gamma_\mu)_\alpha{}^\beta\equiv(-i\sigma_2,\sigma_3,-\sigma_1),
\end{align}
and obey the $SO(1,2)$ Clifford algebra $\{\gamma_\mu,\gamma_\nu\}=2\eta_{\mu\nu}$. The Grassmann bilinear in Eq.~(\ref{superJ}) is thus defined as $\theta\gamma^\nu\bar{\theta}=\theta^\alpha(\gamma^\nu)_\alpha{}^\beta\bar{\theta}_\beta$. One also often uses gamma matrices with two lower or two upper spinorial indices,
\begin{align}\label{GammaMuSymm}
\gamma^\mu_{\alpha\beta}=(-1,\sigma_1,\sigma_3),\hspace{5mm}
\gamma_\mu^{\alpha\beta}=(1,-\sigma_1,-\sigma_3),
\end{align}
which can be obtained from Eq.~(\ref{DefGammaMu}) by raising and lowering the appropriate indices. Importantly, these do \emph{not} satisfy the Clifford algebra, and are real and symmetric. One can use them to write  a Lorentz vector such as $J_\mu$ as a symmetric bispinor $J_{\alpha\beta}$,
\begin{align}\label{bispinors}
J_{\alpha\beta}\equiv\gamma^\mu_{\alpha\beta}J_\mu,\hspace{5mm}
J_\mu=-\frac{1}{2}\gamma_\mu^{\alpha\beta}J_{\alpha\beta},
\end{align}
i.e., a second-rank symmetric tensor in spinorial indices, which has $(2\times 3)/2=3$ independent components, as expected for a Lorentz vector in 2+1 dimensions.

Recent work has determined the general structure of the two-point function of the supercurrent superfield $\mathcal{J}_{\alpha\beta}$ in $\mathcal{N}=2$ superconformal field theories (SCFTs) in 2+1 dimensions~\cite{buchbinder2015},
\begin{align}\label{JJSCFT}
\langle \mathcal{J}_{\alpha\beta}(z_1)\mathcal{J}^{\alpha'\beta'}(z_2)\rangle=c_{\mathcal{N}=2}\frac{\b{x}_{12\alpha}{}^{(\alpha'}\b{x}_{12\beta}{}^{\beta')}}{(\b{x}_{12}{}^2)^3},
\end{align}
where the definition of $\b{x}$ will be given below. On the other hand, the general structure of the two-point functions $\langle J_\mu(x) J_\nu(0)\rangle$ and $\langle T_{\mu\nu}(x)T_{\rho\sigma}(0)\rangle$ in conformal, but not necessarily superconformal, field theories was determined by Osborn and Petkou over two decades ago~\cite{osborn1994},
\begin{align}\label{OsbornPetkou}
\langle J_\mu(x)J_\nu(0)\rangle=C_J\frac{I_{\mu\nu}(x)}{x^4},\hspace{5mm}
\langle T_{\mu\nu}(x)T_{\rho\sigma}(0)\rangle=C_T\frac{I_{\mu\nu,\rho\sigma}(x)}{x^6},
\end{align}
in 2+1 dimensions (we write $x^n\equiv|x|^n$ for simplicity), where the tensors $I_{\mu\nu}(x)$ and $I_{\mu\nu,\rho\sigma}(x)$ are given by
\begin{align}\label{defI}
I_{\mu\nu}(x)\equiv\eta_{\mu\nu}-\frac{2x_\mu x_\nu}{x^2},\hspace{5mm}
I_{\mu\nu,\rho\sigma}(x)\equiv\frac{1}{2}\left(I_{\mu\sigma}(x)I_{\nu\rho}(x)+I_{\mu\rho}(x)I_{\nu\sigma}(x)\right)-\frac{1}{3}\eta_{\mu\nu}\eta_{\rho\sigma}.
\end{align}
Given Eq.~(\ref{superJ}), Eq.~(\ref{JJSCFT}) implies that $C_J$ and $C_T$ are determined by the same universal constant $c_{\mathcal{N}=2}$. By expanding the superspace expression (\ref{JJSCFT}) in Grassmann components, we will determine how $C_J$ and $C_T$ are related.

\subsection{Two-point function of the $U(1)$ current: $C_J$}

By spacetime translation invariance, we can set $z_2$ to zero and $z_1$ to $z$ in Eq.~(\ref{JJSCFT}),
\begin{align}\label{JJx}
\langle \mathcal{J}_{\alpha\beta}(z)\mathcal{J}^{\alpha'\beta'}(0)\rangle=c_{\mathcal{N}=2}\frac{\b{x}_{\alpha}{}^{(\alpha'}\b{x}_{\beta}{}^{\beta')}}{(\b{x}^2)^3},
\end{align}
where we use the notation $A^{(\alpha}B^{\beta)}=\frac{1}{2}(A^\alpha B^\beta+A^\beta B^\alpha)$ for symmetrization. Equation~(5) in the main text follows simply from applying Eq.~(\ref{bispinors}) to Eq.~(\ref{JJx}).  The bispinor $\b{x}^{\alpha\beta}$ is defined as~\cite{buchbinder2015}
\begin{align}
\b{x}^{\alpha\beta}=x^{\alpha\beta}-i\varepsilon^{\alpha\beta}\theta\bar{\theta},
\end{align}
where $\theta\bar{\theta}\equiv\theta^\alpha\bar{\theta}_\alpha$ and $x^{\alpha\beta}$ is the symmetric bispinor corresponding to $x_\mu$. To obtain the two-point function of the $U(1)$ current $J_\mu$, given Eq.~(\ref{superJ}) one must set all $\theta$'s and $\bar{\theta}$'s to zero~\cite{dumitrescu2011} in Eq.~(\ref{JJx}). We have
\begin{align}
s^2\langle J_\mu(x)J_\nu(0)\rangle&=\frac{1}{4}c_{\mathcal{N}=2}\gamma_\mu^{\alpha\beta}(\gamma_\nu)_{\alpha'\beta'}\frac{x_\alpha{}^{(\alpha'}x_\beta{}^{\beta')}}{x^6}
=\frac{1}{4}c_{\mathcal{N}=2}\tr(\gamma_\mu\gamma_\lambda\gamma_\nu\gamma_\rho)\frac{x^\lambda x^\rho}{x^6}
=-\frac{1}{2}c_{\mathcal{N}=2}\frac{I_{\mu\nu}(x)}{x^4},
\end{align}
using the identity $\tr(\gamma_\mu\gamma_\lambda\gamma_\nu\gamma_\rho)=2(\eta_{\mu\lambda}\eta_{\nu\rho}+\eta_{\mu\rho}\eta_{\lambda\nu}
-\eta_{\mu\nu}\eta_{\lambda\rho})$, hence we obtain
\begin{align}
C_J=-\frac{1}{2s^2}c_{\mathcal{N}=2}.
\end{align}

\subsection{Two-point function of the stress tensor: $C_T$}

For the two-point function of the stress tensor, we need to keep only terms that are quadratic in Grassmann variables (with two $\theta$'s and two $\bar{\theta}$'s) on both sides of Eq.~(\ref{JJx}). Using Eq.~(\ref{superJ}), the relevant part of the two-point function of the supercurrent superfield $\mathcal{J}_\mu$ is thus
\begin{align}\label{TTsuper}
\langle \mathcal{J}_\mu(z)\mathcal{J}_\sigma(0)\rangle=4(\theta\gamma^\nu\bar{\theta})(\theta\gamma^\rho\bar{\theta})\langle T_{\nu\mu}(x)T_{\rho\sigma}(0)\rangle+\dotsb
\end{align}
The denominator of Eq.~(\ref{JJx}) is
\begin{align}
(\b{x}^2)^3=\left(-\frac{1}{2}\b{x}^{\alpha\beta}\b{x}_{\alpha\beta}\right)^3=\left(x^2+(\theta\bar{\theta})^2\right)^3=x^6+3x^4(\theta\bar{\theta})^2,
\end{align}
observing that powers of $\theta\bar{\theta}$ higher than two vanish identically because of the Grassmann nature of $\theta$ and $\bar{\theta}$. The compute the numerator, we first observe that
\begin{align}
\b{x}_\alpha{}^{\alpha'}=x_\alpha{}^{\alpha'}-i\delta_\alpha^{\alpha'}\theta\bar{\theta},
\end{align}
and we obtain
\begin{align}
\gamma_\mu^{\alpha\beta}(\gamma_\sigma)_{\alpha'\beta'}\b{x}_\alpha{}^{(\alpha'}\b{x}_\beta{}^{\beta')}=-2x^2I_{\mu\sigma}(x)+2\eta_{\mu\sigma}(\theta\bar{\theta})^2+\dotsb\,,
\end{align}
where $(\dotsb)$ denotes possible terms proportional to $\theta\bar{\theta}$ which do not contribute to the two-point function of the stress tensor. We thus have
\begin{align}\label{TT_RHS}
\langle \mathcal{J}_\mu(z)\mathcal{J}_\sigma(0)\rangle=\frac{c_{\mathcal{N}=2}}{2}\left(\frac{-x^2I_{\mu\sigma}(x)+\eta_{\mu\sigma}(\theta\bar{\theta})^2}{x^6+3x^4(\theta\bar{\theta})^2}\right)+\dotsb
=\frac{c_{\mathcal{N}=2}}{2}\left(\frac{3I_{\mu\sigma}(x)+\eta_{\mu\sigma}}{x^6}\right)(\theta\bar{\theta})^2+\dotsb\,,
\end{align}
where the dots denote all terms not proportional to $(\theta\bar\theta)^2$.
Consider now Eq.~(\ref{TTsuper}). By Lorentz invariance, we must have $(\theta\gamma^\nu\bar{\theta})(\theta\gamma^\rho\bar{\theta})=C\eta^{\nu\rho}(\theta\bar{\theta})^2$ where $C$ is some constant. Setting for instance $\nu=\rho=1$, which implies by Eq.~(\ref{DefGammaMu}) that $\gamma^\nu=\gamma^\rho=\sigma_3$, it is easily shown that $(\theta\sigma_3\bar{\theta})^2=-(\theta\bar{\theta})^2$, and thus
\begin{align}
(\theta\gamma^\nu\bar{\theta})(\theta\gamma^\rho\bar{\theta})=-\eta^{\nu\rho}(\theta\bar{\theta})^2.
\end{align}
Substituting this expression in Eq.~(\ref{TTsuper}), and using Eq.~(\ref{TT_RHS}), we obtain
\begin{align}\label{equationTT}
\eta^{\nu\rho}\langle T_{\nu\mu}(x)T_{\rho\sigma}(0)\rangle=-\frac{1}{8}c_{\mathcal{N}=2}\left(\frac{3I_{\mu\sigma}(x)+\eta_{\mu\sigma}}{x^6}\right),
\end{align}
by equating the coefficients of $(\theta\bar{\theta})^2$ on either side of the equation. To determine the relationship between $C_T$ and $c_{\mathcal{N}=2}$, we compute the left-hand side of Eq.~(\ref{equationTT}) from the general relation Eq.~(\ref{OsbornPetkou}),
\begin{align}
\eta^{\nu\rho}\langle T_{\nu\mu}(x)T_{\rho\sigma}(0)\rangle=\frac{1}{6}C_T\left(\frac{3I_{\mu\sigma}(x)+\eta_{\mu\sigma}}{x^6}\right),
\end{align}
from which we obtain the relation
\begin{align}
C_T=-\frac{3}{4}c_{\mathcal{N}=2}\,.
\end{align}
We thus find that the ratio between $C_J$ and $C_T$ is a universal number: $C_J/C_T=2/3s^2$. To fix the proportionality constant $s$ for the Wess-Zumino theory, we use the fact that the UV fixed point of Eq.~(2) in the main text, the theory of a free boson and a free Dirac fermion, is also a $\mathcal{N}=2$ SCFT in 2+1 dimensions. The coefficients $C_J$ and $C_T$ at this fixed point are simply the sum of the free boson and free Dirac fermion values, which can be computed explicitly~\cite{barnes2005},
\begin{align}
C_J=C_J^\phi+C_J^\psi=10/S_3^2,\hspace{5mm}
C_T=C_T^\phi+C_T^\psi=6/S_3^2, \label{CJCT_UV}
\end{align}
where $S_D\equiv 2\pi^{D/2}/\Gamma(D/2)$. One obtains
\begin{align}
\frac{C_J}{C_T}=\frac{5}{3},\label{ratioCJCT}
\end{align}
and thus $s=\sqrt{2/5}$.

\section{Exact evaluation of the ground-state conductivity $\sigma_\infty$}

In this section we provide the explicit calculation of the ground-state conductivity $\sigma_\infty$ at the semimetal-superconductor
QCP described by the $\mathcal{N}=2$ Wess-Zumino SCFT. We closely follow Nishioka and Yonekura~\cite{nishioka2013},
who gave an integral expression for a quantity that is proportional to $\sigma_\infty$. 
In Ref.~\onlinecite{nishioka2013} this expression was only evaluated numerically, while we here show that this integral, and hence $\sigma_\infty$,
reduces to a simple irrational number. 


Nishioka and Yonekura give an expression for the coefficient $C_T$ of the two-point function of the stress tensor in terms of a quantity called $\tau_{RR}$~\cite{nishioka2013},
\begin{align}
C_T=\frac{3\tau_{RR}}{2\pi^2}.
\end{align}
At the UV fixed point of the Wess-Zumino theory one has $\tau_{RR}=\frac{1}{4}$~\cite{nishioka2013}, in agreement with the value of $C_T$ given in Eq.~(\ref{CJCT_UV}). Using Eq.~(\ref{ratioCJCT}), we thus have
\begin{align}
C_J=\frac{5\tau_{RR}}{2\pi^2}.
\end{align} 
We note that the normalization of the $R$-current in Ref.~\cite{nishioka2013} differs from the one used here~\cite{buchbinder2015}.   
By Fourier transforming the two-point function $\langle JJ\rangle$ in Eq.~(\ref{OsbornPetkou}), 
and using the standard Kubo formula for the conductivity 
\begin{align}
  \sigma(\omega) = -\frac{i}{\omega}\langle J_x(\omega, \vec k=0) J_x(-\omega, \vec k=0) \rangle\,,
\end{align}
we find $\sigma_\infty=\pi^2 C_J/2$, which implies
\begin{align}
  \sigma_\infty = \frac{5}{4}\tau_{RR}\,.
\end{align}

In order to evaluate $\tau_{RR}$, one first considers the partition function of the theory on the compactified spacetime $S_b^3$,
which is a squashed three-sphere. When the squashing parameter $b$ is set to unity, $S_b^3$ reduces to the regular three-sphere.
The (dimensionless) free energy is given by $F(b)=-\log Z_{S_b^3}$, where $Z_{S_b^3}$ is the partition function. 
$\tau_{RR}$ is then obtained by taking the second derivative of $F(b)$ with respect to $b$:
\begin{align} \label{tauRR}
  \tau_{RR} = \frac{2}{\pi^2}\, \text{Re} \left.\frac{\partial^2F}{\partial b^2}\right|_{b=1}\,.
\end{align} 
Heuristically, each $b$-derivative brings down one stress tensor, so that we are left with the two-point function $\langle TT\rangle$. 

The crucial simplification comes because of SUSY, which leads to a powerful method called 
supersymmetric localization that allows the computation 
of the partition function in terms of a simple integral~\cite{imamura2011,imamura2012}. Using Eq.~(\ref{tauRR}) then gives \cite{nishioka2013}
\begin{align}
  \tau_{RR}=\frac{2}{\pi^2}\int_0^\infty dy \left[ \frac{1}{3}\left(\frac{1}{y^2} - \frac{\cosh(2y/3)}{\sinh^2 y} \right) 
  + \frac{[\sinh(2y) -2y]\sinh(2y/3)}{2\sinh^4 y}  \right]\,, \label{localization}
\end{align}
where we have used the fact that the $R$-charge associated with the chiral multiplet of the
interacting Wess-Zumino $\mathcal N=2$ SCFT is $2/3$~\cite{aharony1997}. The fact that Eq.~(\ref{localization}) takes the form of a one-loop
integral but nevertheless describes an interacting conformal QCP follows from SUSY non-renormalization theorems behind the localization
method~\cite{imamura2011,imamura2012}. We emphasize that this integral takes as \emph{input} the exact scaling dimension
of the chiral multiplet (containing $\phi,\psi$).

The part proportional to $1/3$ in Eq.~(\ref{localization}) integrates to:
\begin{align} \label{term1}
  \frac{2}{\pi^2}\int_0^\infty dy\, \frac{1}{3}\left(\frac{1}{y^2} - \frac{\cosh(2y/3)}{\sinh^2 y} \right) = \frac{2}{9\sqrt 3 \pi}\,.
\end{align}
The second term is more subtle. To simplify its evaluation, we slightly deform that part of the integrand: 
\begin{align}
  A(a)= \frac{2}{\pi^2}\int_0^\infty dy\,  \frac{[\sinh(2y) -2y]\sinh(2y/3)}{2\sinh^4 \left(y+\frac{a}{2}\right) }\,,
\end{align}
where we have introduced a shift by $a/2>0$ in the argument of the hyperbolic sine in the denominator. The resulting integral
can be evaluated in closed form (the lengthy answer contains the Lerch transcendent function). Here, we simply give its $a\to 0^+$ limit:  
\begin{align} \label{term2}
  A(0^+)=\frac{64}{243}-\frac{2}{3 \sqrt{3} \pi }\,.
\end{align}
Adding Eqs.~(\ref{term1}) and (\ref{term2}) we obtain:  
\begin{align} \label{tau_RR_final}
  \tau_{RR} =  \frac{4}{243} \left(16-\frac{9 \sqrt{3}}{\pi }\right) \approx 0.1816961307\,,
\end{align}
which agrees with the numerical evaluation of Eq.~(\ref{localization}), given in Ref.~\cite{nishioka2013}.
As an independent check of the SUSY localization calculation described above, a recent highly non-trivial
conformal bootstrap calculation \cite{bobev2015} has yielded $\tau_{RR}=0.18163(8)$, in perfect agreement with the exact result. 
Finally, Eq.~(\ref{tau_RR_final}) leads to the desired result for the ground-state conductivity:    
\begin{align} 
  \sigma_\infty = \frac{5}{4}\tau_{RR} =  \frac{5}{243} \left(16-\frac{9 \sqrt{3}}{\pi }\right) \approx 
0.2271201634 \,.
\end{align}

\section{Three-point function $\langle TJJ\rangle$ of $\mathcal{N}=2$ SCFTs in 2+1 dimensions}

We now consider the three-point function $\langle TJJ\rangle$ of the stress tensor and two $U(1)$ currents. The generic form of this function for $(2+1)$-dimensional CFTs is given by~\cite{osborn1994},
\begin{align}\label{Tjj}
\langle T_{\mu\nu}(x_1)J_\lambda(x_2)J_\rho(x_3)\rangle=\frac{t_{\mu\nu\sigma\tau}(X_{23})\eta^{\sigma\kappa}\eta^{\tau\gamma}I_{\lambda\kappa}(x_{21})I_{\rho\gamma}(x_{31})}{x_{12}^3x_{13}^3x_{23}},
\end{align}
where we define
\begin{align}\label{xijX23}
x_{ij}=x_i-x_j\,, \hspace{5mm}X_{23}=\frac{x_{13}}{x_{13}^2}-\frac{x_{12}}{x_{12}^2}\,.
\end{align}
When the symbol $x_{ij}$ appears raised to an odd power, as in the denominator of Eq.~(\ref{Tjj}), it means $|x_{ij}|=\sqrt{x_{ij}^2}$. The second-rank tensor $I_{\mu\nu}(x)$ is defined in Eq.~(\ref{defI}), and the dimensionless fourth-rank tensor $t_{\mu\nu\sigma\tau}(X)$ is defined as 
\begin{align}
t_{\mu\nu\sigma\tau}(X)=\hat{a}h_{\mu\nu}^1(\hat{X})\eta_{\sigma\tau}+\hat{b}h_{\mu\nu}^1(\hat{X})h_{\sigma\tau}^1(\hat{X})
+\hat{c}h_{\mu\nu\sigma\tau}^2(\hat{X})+\hat{e}h_{\mu\nu\sigma\tau}^3(\hat{X}),
\end{align}
where we define
\begin{align}
h_{\mu\nu}^1(\hat{X})&=\hat{X}_\mu\hat{X}_\nu-\frac{1}{3}\eta_{\mu\nu},\\
h_{\mu\nu\sigma\tau}^2(\hat{X})&=\hat{X}_\mu\hat{X}_\sigma\eta_{\nu\tau}+\hat{X}_\nu\hat{X}_\sigma\eta_{\mu\tau}
+\hat{X}_\mu\hat{X}_\tau\eta_{\nu\sigma}+\hat{X}_\nu\hat{X}_\tau\eta_{\mu\sigma}\nonumber\\
&\hspace{5mm}-\frac{4}{3}\hat{X}_\mu\hat{X}_\nu\eta_{\sigma\tau}-\frac{4}{3}\hat{X}_\sigma\hat{X}_\tau\eta_{\mu\nu}+\frac{4}{9}\eta_{\mu\nu}\eta_{\sigma\tau},\\
h_{\mu\nu\sigma\tau}^3(\hat{X})&=\eta_{\mu\sigma}\eta_{\nu\tau}+\eta_{\mu\tau}\eta_{\nu\sigma}-\frac{2}{3}\eta_{\mu\nu}\eta_{\sigma\tau},
\end{align}
with $\hat{X}_\mu=X_\mu/|X|$. The constants $\hat{a},\hat{b},\hat{c},\hat{e}$ are not linearly independent, as one has the relations
\begin{align}\label{abce}
3\hat{a}-2\hat{b}+2\hat{c}=0,\hspace{5mm}\hat{b}-3\hat{e}=0,
\end{align}
such that the three-point function $\langle TJJ\rangle$ is in general specified by two independent constants.

The form of the three-point function (\ref{JJJSCFT}) simplies tremendously if one considers a collinear frame, i.e., three spacetime points constrained to lie on a straight line: $x_1^\mu=xn^\mu$, $x_2^\mu=yn^\mu$, and $x_3^\mu=zn^\mu$, with $n^\mu n_\mu=1$~\cite{osborn1994}. We also assume for convenience that $x>y>z$. The three-point function is then given by
\begin{align}\label{ATJJ}
\langle T_{\mu\nu}(x_1)J_\lambda(x_2)J_\rho(x_3)\rangle=\frac{\mathcal{A}^{TJJ}_{\mu\nu\lambda\rho}}{(x-y)^3(x-z)^3(y-z)},
\end{align}
where the fourth-rank tensor $\mathcal{A}^{TJJ}_{\mu\nu\lambda\rho}$ is symmetric in both the first and second pair of Lorentz indices.

As seen previously, in $(2+1)$-dimensional $\mathcal{N}=2$ SCFTs both the $U(1)$ current and the stress tensor are part of the supercurrent supermultiplet. Therefore one can extract the three-point function (\ref{Tjj}) from the three-point function of the supercurrent superfield $\mathcal{J}_{\alpha\beta}$, whose general form in those theories has been recently derived~\cite{buchbinder2015},
\begin{align}\label{JJJSCFT}
\langle \mathcal{J}_{\alpha\alpha'}(z_1)\mathcal{J}_{\beta\beta'}(z_2)\mathcal{J}_{\gamma\gamma'}(z_3)\rangle=\frac{\b{x}_{13\alpha\rho}\b{x}_{13\alpha'\rho'}\b{x}_{23\beta\sigma}\b{x}_{23\beta'\sigma'}}{(\b{x}_{13}^2)^3(\b{x}_{23}^2)^3}
H^{\rho\rho'\!,\,\sigma\sigma'}{}_{\gamma\gamma'}(\b{X}_3,\Theta_3),
\end{align}
where the sixth-rank tensor $H$, given in Eq.~(7.44) of Ref.~\onlinecite{buchbinder2015}, is specified by a single independent constant $d_{\mathcal{N}=2}$. Thus $\mathcal{N}=2$ supersymmetry reduces the number of independent constants in the three-point function $\langle TJJ\rangle$ from two to one. Our goal is to determine exactly how $\hat{a},\hat{b},\hat{c},\hat{e}$ are related to $d_{\mathcal{N}=2}$.

To extract Eq.~(\ref{Tjj}) from the superfield expression (\ref{JJJSCFT}), we need only keep the scalar component of $\mathcal{J}_{\beta\beta'}(z_2)$, $\mathcal{J}_{\gamma\gamma'}(z_3)$ and the $\theta\gamma^\mu\bar{\theta}$ component of $\mathcal{J}_{\alpha\alpha'}(z_1)$. Switching from symmetric bispinors to Lorentz vectors, we have
\begin{align}\label{JJJexpanded}
\langle \mathcal{J}_\nu(z_1)\mathcal{J}_\lambda(z_2)\mathcal{J}_\rho(z_3)\rangle=-2s^2(\theta_1\gamma^\mu\bar{\theta}_1)\langle T_{\mu\nu}(x_1)J_\lambda(x_2)J_\rho(x_3)\rangle+\dotsb,
\end{align}
where $\theta_1,\bar{\theta}_1$ are the Grassmann coordinates associated with $z_1$ and the dots represents other components of the superfield three-point function in which we are not interested. Given that $\theta_2,\bar{\theta}_2,\theta_3,\bar{\theta}_3$ do not appear on the right-hand side of Eq.~(\ref{JJJexpanded}), we can set them all to zero in the expansion of Eq.~(\ref{JJJSCFT}) in components. Furthermore only terms quadratic in $\theta_1,\bar{\theta}_1$ need be kept.

The right-hand side of Eq.~(\ref{JJJSCFT}) is expressed in terms of two Grassmann-valued Lorentz vectors $\b{x}_{ij}$ and $\b{X}_3$, and the Lorentz spinor $\Theta_3$, which we must expand in components. The bispinor $\b{x}_{ij}^{\alpha\beta}$ is written as the sum of symmetric and antisymmetric parts~\cite{buchbinder2015},
\begin{align}
\b{x}_{ij}^{\alpha\beta}=\tilde{x}_{ij}^{\alpha\beta}+\frac{i}{2}\varepsilon^{\alpha\beta}\theta_{ijI}^\gamma\theta_{ijI\gamma},
\end{align}
where $\theta_{ijI}^\alpha\equiv\theta_{iI}^\alpha-\theta_{jI}^\alpha$, $I=1,2$, are differences of real Grassmann coordinates. The complex coordinates $\theta_i^\alpha,\bar{\theta}_i^\alpha$ are given in terms of the latter as
\begin{align}
\theta_i^\alpha=\frac{1}{\sqrt{2}}\left(\theta_{i1}^\alpha+i\theta_{i2}^\alpha\right),\hspace{5mm}
\bar{\theta}_i^\alpha=\frac{1}{\sqrt{2}}\left(\theta_{i1}^\alpha-i\theta_{i2}^\alpha\right),
\end{align}
hence $\theta_{ijI}^\alpha\theta_{ijI\alpha}=2\theta_{ij}^\alpha\bar{\theta}_{ij\alpha}=2\theta_{ij}\bar{\theta}_{ij}$, and we can write
\begin{align}
\b{x}_{ij}^{\alpha\beta}=\tilde{x}_{ij}^{\alpha\beta}+i\varepsilon^{\alpha\beta}\theta_{ij}\bar{\theta}_{ij}.
\end{align}
The symmetric part is defined as
\begin{align}\label{xijSymmPart}
\tilde{x}_{ij}^{\alpha\beta}=x_{ij}^{\alpha\beta}+2i\theta_{iI}^{(\alpha}\theta_{jI}^{\beta)},
\end{align}
where $x_{ij}^{\alpha\beta}$ is the symmetric bispinor associated with $x_{ij}$ defined in Eq.~(\ref{xijX23}). Since $i\neq j$, the second term on the right-hand side of Eq.~(\ref{xijSymmPart}) necessarily involves Grassmann coordinates other than $\theta_1,\bar{\theta}_1$, and we can write $\tilde{x}_{ij}^{\alpha\beta}=x_{ij}^{\alpha\beta}$. Therefore, for our purposes $\b{x}_{ij}^{\alpha\beta}=x_{ij}^{\alpha\beta}+i\varepsilon^{\alpha\beta}\theta_{ij}\bar{\theta}_{ij}$. Furthermore, since $\theta_2,\bar{\theta}_2,\theta_3,\bar{\theta}_3$ are set to zero, we have
\begin{align}
\b{x}_{12}^{\alpha\beta}&=x_{12}^{\alpha\beta}+i\varepsilon^{\alpha\beta}\theta_1\bar{\theta}_1,
\hspace{5mm} \b{x}_{12\alpha\beta}=x_{12\alpha\beta}-i\varepsilon_{\alpha\beta}\theta_1\bar{\theta}_1,\label{x12ab}\\
\b{x}_{13}^{\alpha\beta}&=x_{13}^{\alpha\beta}+i\varepsilon^{\alpha\beta}\theta_1\bar{\theta}_1, 
\hspace{5mm} \b{x}_{13\alpha\beta}=x_{13\alpha\beta}-i\varepsilon_{\alpha\beta}\theta_1\bar{\theta}_1,\\
\b{x}_{23}^{\alpha\beta}&=x_{23}^{\alpha\beta}, 
\hspace{24.5mm} \b{x}_{23\alpha\beta}=x_{23\alpha\beta}.\label{x23ab}
\end{align}
Since $x_{ij}^{\alpha\beta},x_{ij\alpha\beta}$ are symmetric in spinor indices and $\varepsilon^{\alpha\beta},\varepsilon_{\alpha\beta}$ antisymmetric, the square of $\b{x}_{ij}$ is given by $\b{x}_{ij}^2\equiv-\frac{1}{2}\b{x}_{ij}^{\alpha\beta}\b{x}_{ij\alpha\beta}=x_{ij}^2+\frac{1}{4}(\theta_1\bar{\theta}_1)^2=x_{ij}^2$. The bispinor $\b{X}_{3\alpha\beta}$ is defined as the matrix elements of~\cite{buchbinder2015}
\begin{align}
\check{\b{X}}_3=-\frac{\check{\b{x}}_{13}^T\hat{\b{x}}_{12}\check{\b{x}}_{32}^T}{\b{x}_{13}^2\b{x}_{32}^2},
\end{align}
where $\hat{\b{m}}$ denotes a matrix with two upper spinorial indices and $\check{\b{m}}$ a matrix with two lower spinorial indices. We thus have
\begin{align}\label{X3ab}
\b{X}_{3\alpha\beta}=-\frac{\b{x}_{13\alpha'\alpha}\b{x}_{12}^{\alpha'\beta'}\b{x}_{32\beta\beta'}}{\b{x}_{13}^2\b{x}_{32}^2}=\frac{x_{32\beta\beta'}}{x_{13}^2x_{32}^2}\left(x_{13\alpha}{}^{\alpha'}x_{12\alpha'}{}^{\beta'}+ix_{23\alpha}{}^{\beta'}\theta_1\bar{\theta}_1\right),
\end{align}
to the desired order. Finally, $\Theta_{3I\alpha}$ is defined as the matrix elements of~\cite{buchbinder2015}
\begin{align}
\hat{\Theta}_3=-\frac{\check{\b{x}}_{13}^T\hat{\theta}_{31}}{\b{x}_{13}^2}+\frac{\check{\b{x}}_{23}^T\hat{\theta}_{32}}{\b{x}_{23}^2},
\end{align}
where $\hat{\theta}_{ij}$ is a matrix with matrix elements $\theta_{ijI}^\alpha=\theta_{iI}^\alpha-\theta_{jI}^\alpha$. To the desired order, we obtain
\begin{align}\label{Theta3Ia}
\Theta_{3I\alpha}=\frac{x_{13\alpha\beta}\theta_{1I}^\beta}{x_{13}^2}.
\end{align}

We now evaluate Eq.~(\ref{JJJSCFT}) in the collinear frame discussed earlier. In the collinear frame, Eq.~(\ref{x12ab})-(\ref{x23ab}) simplify to
\begin{align}
\b{x}_{12\alpha\beta}&=(x-y)n_{\alpha\beta}-i\varepsilon_{\alpha\beta}\theta_1\bar{\theta}_1,\\
\b{x}_{13\alpha\beta}&=(x-z)n_{\alpha\beta}-i\varepsilon_{\alpha\beta}\theta_1\bar{\theta}_1,\\
\b{x}_{23\alpha\beta}&=(y-z)n_{\alpha\beta},
\end{align}
where $n_{\alpha\beta}$ is the symmetric bispinor associated with $n^\mu$, and Eq.~(\ref{X3ab}), (\ref{Theta3Ia}) become
\begin{align}\label{X3abColl}
\b{X}_{3\alpha\beta}&=-\frac{x-y}{(x-z)(y-z)}n_{\alpha\beta}+\frac{i}{(x-z)^2}\varepsilon_{\alpha\beta}\theta_1\bar{\theta}_1,\\
\Theta_{3I\alpha}&=\frac{1}{x-z}n_{\alpha\beta}\theta_{1I}^\beta.
\end{align}
The sixth-rank tensor $H$ in Eq.~(\ref{JJJSCFT}) is given by~\cite{buchbinder2015}
\begin{align}\label{H}
H^{\alpha\alpha',\beta\beta',\gamma\gamma'}(\b{X}_3,\Theta_3)&=id_{\mathcal{N}=2}\biggl[
\frac{2}{\b{X}_3^3}\left(\varepsilon^{\alpha(\beta}\varepsilon^{\beta')\alpha'}\Xi_3^{\gamma\gamma'}
+\varepsilon^{\alpha(\gamma}\varepsilon^{\gamma')\alpha'}\Xi_3^{\beta\beta'}
+\varepsilon^{\beta(\gamma}\varepsilon^{\gamma')\beta'}\Xi_3^{\alpha\alpha'}\right)\nonumber\\
&+\frac{1}{\b{X}_3^5}\left(3\b{X}_3^{\alpha\alpha'}\b{X}_3^{\gamma\gamma'}\Xi_3^{\beta\beta'}
+3\b{X}_3^{\beta\beta'}\b{X}_3^{\gamma\gamma'}\Xi_3^{\alpha\alpha'}-5\b{X}_3^{\alpha\alpha'}\b{X}_3^{\beta\beta'}\Xi_3^{\gamma\gamma'}\right)\nonumber\\
&+\frac{1}{\b{X}_3^5}\left(5\varepsilon^{\alpha(\gamma}\varepsilon^{\gamma')\alpha'}\b{X}_3^{\beta\beta'}
+5\varepsilon^{\beta(\gamma}\varepsilon^{\gamma')\beta'}\b{X}_3^{\alpha\alpha'}
-3\varepsilon^{\alpha(\beta}\varepsilon^{\beta')\alpha'}\b{X}_3^{\gamma\gamma'}\right)\b{X}_3^{\delta\delta'}\Xi_{3\delta\delta'}\nonumber\\
&+\frac{5}{2}\frac{1}{\b{X}_3^7}\b{X}_3^{\alpha\alpha'}\b{X}_3^{\beta\beta'}\b{X}_3^{\gamma\gamma'}\b{X}_3^{\delta\delta'}\Xi_{3\delta\delta'}\biggr],
\end{align}
where we define
\begin{align}
\Xi_{3\alpha\alpha'}=\varepsilon_{IJ}\Theta_{3I\alpha}\Theta_{3J\alpha'}.
\end{align}
In the collinear frame, we find
\begin{align}
\Xi_{3\alpha\alpha'}=\frac{2i}{(x-z)^2}n_{\alpha\beta}n_{\alpha'\beta'}\theta_1^{(\beta}\bar{\theta}_1^{\beta')}.
\end{align}
Since each term in Eq.~(\ref{H}) contains $\Xi_3$, which is quadratic in the Grassmann coordinates $\theta_1,\bar{\theta}_1$, we can ignore the Grassmann part of $\b{X}_3$ in Eq.~(\ref{X3abColl}). We therefore have
\begin{align}
\b{X}_{3\alpha\beta}=-\frac{x-y}{(x-z)(y-z)}n_{\alpha\beta},\hspace{5mm}|\b{X}_3|=\sqrt{-\half\b{X}_3^{\alpha\beta}\b{X}_{3\alpha\beta}}=\frac{x-y}{(x-z)(y-z)}.
\end{align}
Likewise, since $H$ is quadratic in Grassmann coordinates we can neglect the Grassmann part of the $\b{x}_{ij\alpha\beta}$ factors in front of $H$ in Eq.~(\ref{JJJSCFT}). Using the latter equation, we have
\begin{align}\label{ContractedTJJ}
s^2(\theta_1\gamma^\mu\bar{\theta}_1)\langle T_{\mu\nu}(x_1)J_\lambda(x_2)J_\rho(x_3)\rangle=\frac{1}{16(x-z)^6(y-z)^6}\Delta_{\nu\lambda\rho}(x,y,z),
\end{align}
where
\begin{align}
\Delta_{\nu\lambda\rho}(x,y,z)&=\gamma_\nu^{\alpha\alpha'}
\gamma_\lambda^{\beta\beta'}
\gamma_\rho^{\gamma\gamma'}
x_{13\alpha\tau}x_{13\alpha'\tau'}x_{23\beta\sigma}
x_{23\beta'\sigma'}H^{\tau\tau',\sigma\sigma'}{}_{\gamma\gamma'}(\b{X}_3,\Theta_3)\nonumber\\
&=(x-z)^2(y-z)^2(n\gamma_\nu n)_{\alpha\alpha'}(n\gamma_\lambda n)_{\beta\beta'}(\gamma_\rho)_{\gamma\gamma'}
H^{\alpha\alpha',\beta\beta',\gamma\gamma'}(\b{X}_3,\Theta_3).
\end{align}

For simplicity we will focus only on terms in Eq.~(\ref{ContractedTJJ}) that are proportional to $\theta_1\gamma_\lambda\bar{\theta}_1$, as this turns out to be sufficient to relate $\hat{a},\hat{b},\hat{c},\hat{e}$ to $d_{\mathcal{N}=2}$. Since on the right-hand side of Eq.~(\ref{ContractedTJJ}) $\theta_1,\bar{\theta}_1$ only appear in $\Xi_3$, and since the index $\lambda$ appears in the combination $(n\gamma_\lambda n)_{\beta\beta'}$, the only terms in Eq.~(\ref{H}) that can generate $\theta_1\gamma_\lambda\bar{\theta}_1$ are those that contain $\Xi_3^{\beta\beta'}$. Using
\begin{align}
(n\gamma_\lambda n)_{\beta\beta'}\Xi_3^{\beta\beta'}&=\frac{2i}{(x-z)^2}\theta_1\gamma_\lambda\bar{\theta}_1,\\
(n\gamma_\nu n)_{\alpha\alpha'}(\gamma_\rho)_{\gamma\gamma'}\varepsilon^{\alpha(\gamma}\varepsilon^{\gamma')\alpha'}&=-2\eta_{\nu\rho}+4n_\nu n_\rho,\\
(n\gamma_\nu n)_{\alpha\alpha'}(\gamma_\rho)_{\gamma\gamma'}\b{X}_3^{\alpha\alpha'}\b{X}_3^{\gamma\gamma'}&=\frac{4(x-y)^2}{(x-z)^2(y-z)^2}n_\nu n_\rho,
\end{align}
as well as Eq.~(\ref{ContractedTJJ}) and (\ref{ATJJ}), we obtain
\begin{align}\label{ContractedATJJ}
s^2(\theta_1\gamma^\mu\bar{\theta}_1)\mathcal{A}^{TJJ}_{\mu\nu\lambda\rho}=\frac{1}{2}d_{\mathcal{N}=2}\left(\eta_{\nu\rho}-5n_\nu n_\rho\right)\theta_1\gamma_\lambda\bar{\theta}_1+\ldots
\end{align}
On the other hand, we can directly calculate the left-hand side of Eq.~(\ref{ContractedATJJ}) for a general CFT in 2+1 dimensions from Eq.~(\ref{Tjj}) and (\ref{ATJJ}). We obtain
\begin{align}\label{ATJJ_CFT}
s^2(\theta_1\gamma^\mu\bar{\theta}_1)\mathcal{A}^{TJJ}_{\mu\nu\lambda\rho}=\frac{s^2}{3}\left(\hat{b}\eta_{\nu\rho}-(2\hat{b}+3\hat{c})n_\nu n_\rho\right)\theta_1\gamma_\lambda\bar{\theta}_1+\ldots,
\end{align}
where we have used the relations (\ref{abce}) to eliminate $\hat{a}$ and $\hat{e}$ in favor of $\hat{b}$ and $\hat{c}$. Comparing Eq.~(\ref{ContractedATJJ}) and (\ref{ATJJ_CFT}), we obtain
\begin{align}
\hat{b}=\hat{c}=\frac{3}{2s^2}d_{\mathcal{N}=2},
\end{align}
which implies using the relations (\ref{abce}) that
\begin{align}
\hat{a}=0,\hspace{5mm}\hat{e}=\frac{1}{2s^2}d_{\mathcal{N}=2}.
\end{align}
Refs.~\cite{hofman08,ajay,chowdhury2013} define a constant $\gamma$, which is a function of the constants $\hat{a},\hat{b},\hat{c},\hat{e}$ of the CFT. 
This constant was shown to be subject to certain bounds~\cite{hofman08,ajay}. 
In 2+1 dimensions, $\gamma$ is given by
\begin{align}
\gamma=\frac{\hat{b}-\hat{c}}{4\hat{b}+12\hat{c}}. 
\end{align}
We thus find that all (2+1)D CFTs with $\mathcal{N}=2$ superconformal invariance will have a vanishing value of $\gamma$ (as long as the $U(1)$ current is proportional to the $R$-current):
\begin{align}
\gamma_{\mathcal{N}=2}=0\,.
\end{align}

\bibliography{susy}